\documentclass[journal]{IEEEtran}
\usepackage{epsfig,graphics,subfigure,psfrag,amsmath,cases}
\usepackage{latexsym,amssymb,amsmath,epsfig,subfigure,algorithm,amsthm}
\usepackage{algorithmic}
\usepackage{color}
\usepackage{url}
\usepackage{scrtime}
\usepackage{bm}
\usepackage{graphicx}
\usepackage{subfigure}
\usepackage{bbding}
\usepackage{multicol}
\usepackage{graphics}
\usepackage{amsthm}
\usepackage{array}

\author{Yuanxin Cai,~\IEEEmembership{Student Member,~IEEE,} Zhiqiang Wei,~\IEEEmembership{Member,~IEEE,} Ruide Li, \\
Derrick Wing Kwan Ng,~\IEEEmembership{Senior Member,~IEEE,}\thanks{ D. W. K. Ng is supported by funding from the UNSW Digital Grid Futures Institute, UNSW, Sydney, under a cross-disciplinary fund scheme and by the Australian Research Council's Discovery Project (DP190101363). }
 and Jinhong Yuan~\IEEEmembership{Fellow,~IEEE}
\thanks{J. Yuan is supported in part by the Australian Research Council Discovery Projects under Grant d DP190101363 and in part by the Linkage Project under Grant LP170101196.}
\thanks{
Y. Cai, Z. Wei, D. W. K. Ng, and J. Yuan are with the School of Electrical Engineering and Telecommunications, the University of New South Wales, Australia;
R. Li is with the School of Information and Electronics, Beijing Institute of Technology, China.
(e-mail: \{yuanxin.cai, zhiqiang.wei, w.k.ng, j.yuan\}@unsw.edu.au, taiyuanlaide@163.com). \emph{Corresponding author: Zhiqiang Wei}.
This paper has been presented in part at IEEE WCNC 2019 \cite{cai2019energy}.}
}

\title{Joint Trajectory and Resource Allocation Design for Energy-Efficient Secure UAV Communication Systems}

{\newtheorem{Thm}{Theorem}}

\newtheorem{Remark}{Remark}

\begin{document}
\maketitle
\begin{abstract}

  In this paper, we study the trajectory and resource allocation design for downlink energy-efficient secure unmanned aerial vehicle (UAV) communication systems, where an information UAV assisted by a multi-antenna jammer UAV serves multiple ground users in the existence of multiple ground eavesdroppers.
  The resource allocation strategy and the trajectory of the information UAV, and the jamming policy of the jammer UAV are jointly optimized for maximizing the system energy efficiency.
  The joint design is formulated as a non-convex optimization problem taking into account the quality of service (QoS) requirement, the security constraint, and the imperfect channel state information (CSI) of the eavesdroppers.
  The formulated problem is generally intractable.
  As a compromise approach, the problem is divided into two subproblems which facilitates the design of a low-complexity suboptimal algorithm based on alternating optimization approach.
  Simulation results illustrate that the proposed algorithm converges within a small number of iterations and demonstrate some interesting insights:
  (1) the introduction of a jammer UAV facilitates a highly flexible trajectory design of the information UAV which is critical to improving the system energy efficiency;
  (2) by exploiting the spatial degrees of freedom brought by the multi-antenna jammer UAV, our proposed design can focus the artificial noise on eavesdroppers offering a strong security mean to the system.


\end{abstract}

\section{Introduction}

\IEEEPARstart{R}{ecently}, there are rapid growth of expectations on future wireless networks, e.g., ultra-high data rates, low latency, and massive connectivity, etc., \cite{wong2017key}, which pose enormous challenges on the existing wireless communications and related facilities.
Although existing technologies, e.g., multiple-input multiple-output (MIMO), offer a temporary solution to the problems \cite{8632710}\nocite{8535085,8198807}--\cite{8529214}, providing high-data-rate communications in emergencies and important scenarios, such as natural disasters and overloaded traffic demand, remains challenging.
{Fortunately, unmanned aerial vehicles (UAVs)-assisted communication systems serve as a viable solution \cite{8675384}\nocite{opportunities_and_challenges,8660516}--\cite{8918497}, which relax the limitation of traditional wireless communications on the physical layer.}
In particular, by exploiting the high flexibility and mobility of UAVs, the performance of the communication systems can be improved by moving UAVs close to the desired users.
Besides, in practice, UAVs offer a higher probability to establish a strong line-of-sight (LoS) wireless channels between UAVs and ground terminals compared to conventional terrestrial communication systems.
Therefore, in recent years, there are several exciting and practical applications of UAV proposed in academia, such as mobile base stations \cite{DWK_2018_solar_power_comm,8644071}, mobile relays \cite{relays}, and mobile data collections \cite{sensor_network}, etc.

In practice, although UAV-based communications enjoy various advantages, some technical problems need to be solved to unlock the promised performance gains.
Firstly, stringent power limitation is one of the bottlenecks for enabling efficient UAV communications.
In fact, the energy storage of onboard battery of a UAV is usually small due to the size and weight restrictions of the UAV.
Besides, the power consumptions of flight and communication depends on the UAV's trajectory and velocity.
As a result, energy-efficient UAV has drawn significant research interests in the literature.
For example, the authors in \cite{sensor_network} studied the energy efficiency maximization for wireless sensor networks via jointly optimizing the weak up schedule of sensor nodes and UAV's trajectory.
Yet, the flight power consumption of the system was not considered which contributes a significant portion of total system power consumption.
Besides, the UAV trajectory design was developed to optimize the system energy efficiency in \cite{EE_fixed_wing}.
However, the joint investigation of variable speed and transmit power allocation strategy for communications was not conducted which plays an important role for the design of energy-efficient UAV systems.
On the other hand, in order to support simultaneous energy-efficient multi-user communications, orthogonal frequency division multiple access (OFDMA) is an ideal candidate, as it has been commonly adopted in various conventional communication systems \cite{ng_EE,Ng_EE_limi_backhaul,Ng_EE_full_duplex}.
In particular, OFDMA provides a high flexibility in resource allocation for exploiting multi-user diversity to improve the system energy efficiency.
In \cite{UAV_OFDMA}, OFDMA was adopted for UAV communication systems and a joint trajectory and resource allocation design was proposed to maximize the minimum data rate.
However, an energy-efficient design for UAV-OFDMA system has not been reported in the literature, yet.

{Secondly, since the LoS dominated channels between a UAV and ground nodes are susceptible to potential eavesdropping \cite{opportunities_and_challenges,8411465}, guaranteeing communication security is a challenging task for UAV communication systems.}
Thus, there is an emerging need for designing secure UAV-based communication.
For instance, the authors in \cite{Zhang2018Securing} proposed a joint power allocation and trajectory design to maximize the secrecy rate in both uplink and downlink systems.
In \cite{secrecy_EE}, secure energy efficiency maximization for UAV-based relaying systems was studied. However, both works only considered the case of single-user and the proposed designs in \cite{Zhang2018Securing,secrecy_EE} are not applicable to the case of multiple users.
Besides, the availability of the eavesdropper location was assumed in \cite{Zhang2018Securing,secrecy_EE}, which is generally over optimistic.
Although \cite{cui2018robust} studied the resource allocation design for secure UAV systems by taking into account the imperfect channel state information (CSI) of an eavesdropper, the energy efficiency of such systems is still an unknown. Besides, a robust trajectory and resource allocation design for energy-efficient secure UAV communication systems considering the uncertainty of eavesdropper's location has not been investigated.
{Furthermore, although deploying a single UAV in the system was demonstrated to offer some advantages for wireless communications \cite{cai2019energy,7875081}, the performance of single UAV communication systems can be unsatisfactory due to the stringent requirement on secure communication.}
Thus, with the assistance of a jammer UAV, the authors in \cite{8589002,8643815,8408554} proposed a cooperative jamming scheme for secure UAV communications by jointly optimizing power allocation and trajectories.
Yet, since the jammer UAV is only equipped with a single-antenna in these systems, the direction of artificial noise cannot be controlled properly which also causes strong interference to legitimate users due to the existence of strong LoS paths.
Therefore, we propose to employ multiple antennas at the jammer UAV to focus the artificial noise to degrade the channel quality of eavesdroppers as well as to mitigate the interference upon legitimate users.
However, designing a cooperative jamming policy with a multi-antenna jammer UAV is very challenging and remains to be explored.

In this paper, we study the joint trajectory, resource allocation, and jamming policy design for energy-efficient secure UAV-OFDMA communication systems.
In particular, an information UAV provides energy-efficient secure communication for multiple legitimate users adopting OFDMA in the existence of multiple eavesdroppers, with the assistance of a multiple-antenna jammer UAV patrolling with a fixed trajectory.
The joint design is formulated as a non-convex optimization problem to maximize the system energy efficiency taking into account the maximum tolerable leakage signal-to-interference-plus-noise ratio (SINR) to eavesdroppers and the minimum individual user data rate requirement.
Since the formulated problem is non-convex which is generally intractable, we propose an iterative algorithm to achieve a suboptimal solution of the formulated problem.
To this end, we first divide the formulated problem into two sub-problems and solve them alternatively via alternating optimization.
In each iteration, a suboptimal solution can be achieved by employing successive convex approximation (SCA) and the Dinkelbach's method with fast convergence.

The remainder of this paper is organized as follows.
In Section II, we introduce the proposed downlink UAV-enabled communication system model.
The optimization problem formulation is provided in Section III.
In Section IV, we propose an efficient iterative algorithm based on SCA and the Dinkelbach's method to obtain a suboptimal solution of the formulated problem.
Section V provides some numerical results to evaluate the performance of the proposed algorithm.
Finally, the paper is concluded in Section VI.

\section{System Model}

\subsection{Notation}

$\mathbb{R}^{M \times N}$ and $\mathbb{C}^{M \times N}$ denote the the space of an $M \times N$ matrix with real and complex values, respectively.
$\mathbb{H}^{M}$ is an $M \times M$ complex hermitian matrix.
$\|\cdot\|$ denotes the vector norm and $\mathbf{I}_n$ represents an $n \times n$ identity matrix.
$\{\mathbf{A}\}_{r,c}$ denote the element at the $r$-th row and $c$-th column of the matrix $\mathbf{A}$.
For a square-matrix $\mathbf{X}$, $\mathbf{X} \succeq \mathbf{0}$ denotes that $\mathbf{X}$ is a positive semi-definite matrix and $\text{Tr}(\mathbf{X})$ is the trace of the matrix.
$\mathbf{X}^{\mathrm H}$ and $\text{Rank}(\mathbf{X})$ represent the conjugate transpose and the rank of matrix $\mathbf{X}$, respectively.
$\mathbf{X} \otimes \mathbf{Y}$ represents the Kronecker product of two matrices $\mathbf{X}$ and $\mathbf{Y}$.
The distribution of a circularly symmetric complex Gaussian (CSCG) vector with mean vector $\mathbf{x}$ and covariance matrix $\bm \Sigma$ is denoted by $\mathcal{CN}(\mathbf{x},\bm \Sigma)$, and $\sim$ means ``distributed as''.
$\mathcal{O}(\cdot)$ denotes the big-O notation.

\subsection{Signal Model}

\begin{figure}[t]
  \centering
  \includegraphics[width=3.5 in]{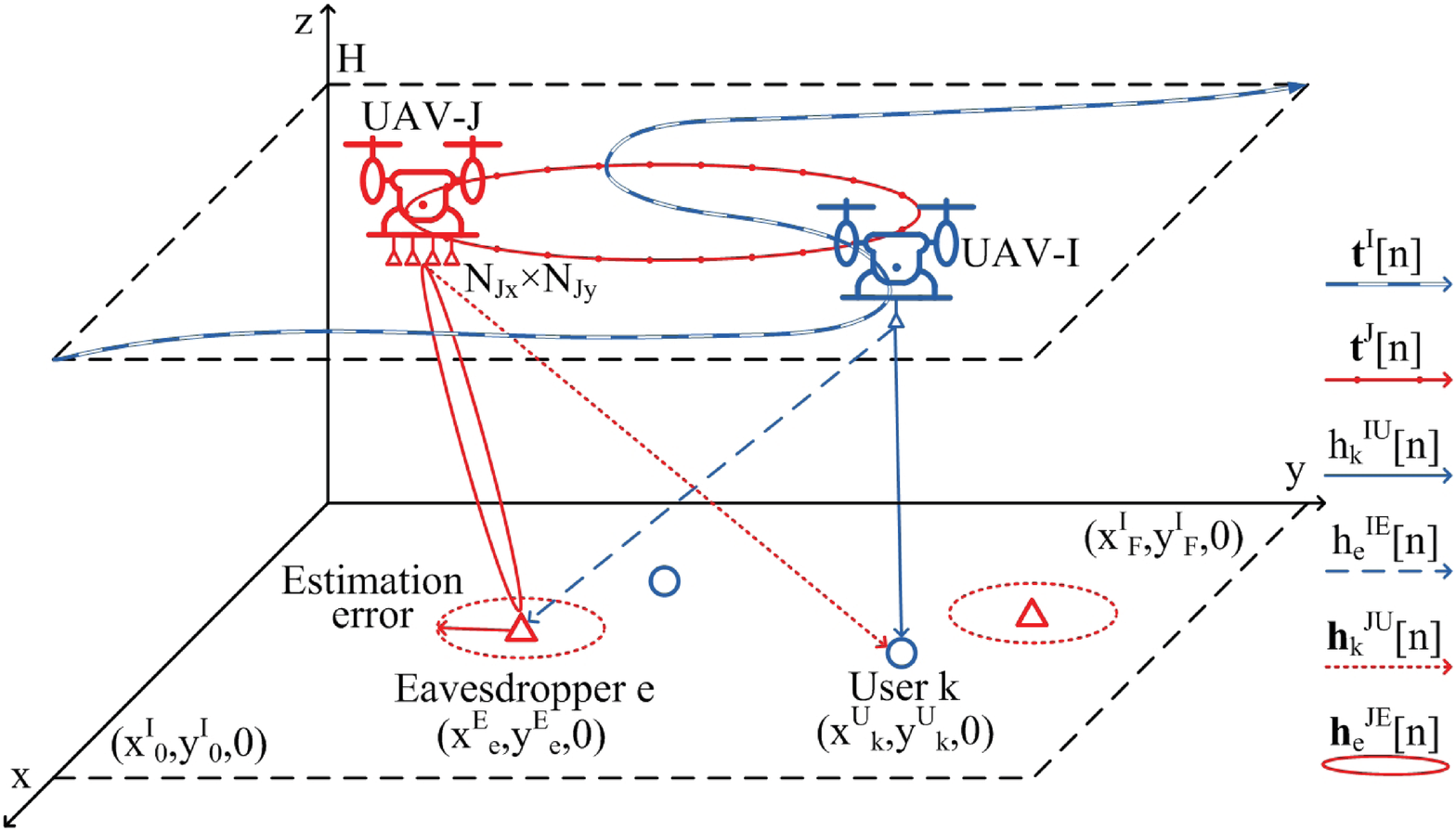} 
  \caption{A UAV-OFDMA system with a multi-antenna jammer UAV, multiple legitimate users, and multiple potential eavesdroppers. The dotted circles denote the uncertainty of the eavesdroppers.}
  \label{system_model}
\end{figure}

A UAV-based OFDMA\footnote{In this paper, we consider a more general problem formulation where user scheduling is performed in subcarrier-level.
This study is applicable to the special case where resource allocation is performed in resource block levels.} communication system is considered which consists of a UAV serving as an information transmitter, $K$ legitimate users, and another UAV serving as a jammer to combat $E$ non-cooperative potential eavesdroppers, as shown in Figure \ref{system_model}.
The information UAV, the legitimate users, and the potential eavesdroppers are single-antenna devices.
On the other hand, we assume that the jammer UAV is equipped with $N_{\mathrm J} = N_{\mathrm Jx} \times N_{\mathrm Jy}$ antennas such that $N_{\mathrm J} > E$.
Besides, artificial noise is generated from the jammer UAV and is steered towards eavesdroppers for ensuring communication security.
To facilitate the system design and simplicity, the jammer UAV patrols the service area with a fixed trajectory and a constant flight velocity\footnote{In this paper, we assume that the jammer UAV has a fixed trajectory and a constant flight velocity to simplify the design of resource allocation.
Note that the proposed framework can achieve a superior performance compared to existing designs, e.g., \cite{7875081,cui2018robust}, as will be verified in the simulation section.
Optimizing jammer UAV's trajectory is an interesting but challenging work and will be considered in our future study.}.
Note that although the jammer UAV cruises with a defined trajectory, it can generate focused artificial noise to interference the eavesdroppers via exploiting the spatial degrees of freedom brought by the multiple antennas.
We assume that the total bandwidth and the time duration of the system are divided equally into $N_{\mathrm{F}}$ subcarriers and $N$ time slots, respectively.
Besides, in the system, we assume that the information UAV and the jammer UAV operate at a constant altitude\footnote{We note that since the channel between the UAV and the ground terminals are LoS dominated \cite{secrecy_EE,cui2018robust,8643815}, the UAVs would fly at the lowest allowable flight altitude to obtain a higher channel gain for maximizing the system energy efficiency.
Thus, we consider a fixed UAVs' flight altitude of $H = 100$ m.} $H$ and all the ground nodes, i.e., legitimate users and eavesdroppers, are fixed during $N$ time slots.
To facilitate secure communication, artificial noise is generated $\mathbf{z}_i^{\mathrm J}[n] \in \mathbb{C}^{N_{\mathrm J} \times 1}$ on subcarrier $i\in\{1,\ldots,N_{\mathrm F}\}$ at time slot $n\in\{1,\ldots,N\}$ by the jammer UAV.
Note that the duration of each time slot $n$ is denoted by $\tau$.
Furthermore, we assume that $\mathbf{z}_i^{\mathrm J}[n]$ can be modeled by a complex Gaussian random vector: 
\begin{eqnarray}
\mathbf{z}_i^{\mathrm J}[n] \thicksim \mathcal{CN}(\mathbf{0},\mathbf{Z}_i^{\mathrm J}[n]),
\end{eqnarray}
where $\mathbf{Z}_i^{\mathrm J}[n] \in \mathbb{H}^{N_{\mathrm J}}$ with $\mathbf{Z}_i^{\mathrm J}[n] \succeq \mathbf{0}$ represents the covariance matrix of the artificial noise on subcarrier $i$ at time slot $n$.
The artificial noise signal $\mathbf{z}_i^{\mathrm J}[n]$ is unknown to both the legitimate receivers and the potential eavesdroppers.
{We introduce a multi-antenna jammer UAV to assist the UAV-based communication system to guarantee secure communication.
Although the additional artificial noise generated by the jammer UAV may cause interference to legitimate ground users, the artificial noise is optimized and mainly focused on the eavesdroppers.
If the jamming does not improve the system performance, the proposed optimization framework will set $\mathbf{Z}_i^{\mathrm J}[n] = \mathbf{0}$ automatically to shut down the artificial noise transmission.}
{In the considered system, the air-to-ground channel is dominated by LoS links with a reasonable flight height and all size \cite{8337920,colpaert2018aerial}.
To simplify the design in the sequel, we assume that the channel is modeled by pure LoS links as commonly adopted in the literature, e.g., \cite{sensor_network,EE_fixed_wing,Zhang2018Securing,secrecy_EE}.}
As the UAV communication channel is dominated by the LoS links\footnote{Based on field measurements \cite{8337920,colpaert2018aerial}, the air-to-ground links between the UAVs and the ground terminals are LoS channels in rural areas when the flight altitude of a UAV is 100 meters and the length of side of the service area is 500 meters.
Besides, the adopted LoS model can facilitate the design of resource allocation and trajectory in the sequel.}, the CSI between each node and each UAV can be determined by its location \cite{Zhang2018Securing,secrecy_EE,cui2018robust,8589002,8643815}.
Besides, the desired ground node users perform handshaking with the system regularly such that accurate location information is available for resource allocation design.
In contrast, since potential eavesdroppers are usually less active in the systems, we assume that only the estimations of their locations are available.
Thus, the distances between the information UAV and user $k\in\{1,\ldots,K\}$ as well as the jammer UAV\footnote{We assume that all the antennas have roughly the same distance between the jammer UAV and user $k$.
In fact, this assumption generally holds as antenna separation at the jammer is generally much shorter compared to the distance between the jammer UAV and ground users.} and user $k$ at time slot $n$ are given by
\begin{align}
d_k^{\mathrm{IU}}[n] &= \sqrt{\|\mathbf{t}_k^{\mathrm U} - \mathbf{t}^{\mathrm I}[n]\|^2 + H^2} \,\, \text{and} \\
d_k^{\mathrm{JU}}[n] &= \sqrt{\|\mathbf{t}_k^{\mathrm U} - \mathbf{t}^{\mathrm J}[n]\|^2 + H^2} ,
\end{align}
respectively.
$\mathbf{t}_k^{\mathrm U} = [x_k^{\mathrm U},y_k^{\mathrm U}]^{\mathrm{T}} \in \mathbb{R}^{2 \times 1} $ represents the location of ground user $k$, $\mathbf{t}^{\mathrm I}[n] = [x^{\mathrm I}[n], y^{\mathrm I}[n]]^{\mathrm{T}} \in \mathbb{R}^{2 \times 1}$ and $\mathbf{t}^{\mathrm J}[n] = [x^{\mathrm J}[n], y^{\mathrm J}[n]]^{\mathrm{T}} \in \mathbb{R}^{2 \times 1}$ represent the horizontal location of the information UAV and the jammer UAV at time slot $n$, respectively.
Similarly, the distance between the information UAV and potential eavesdropper $e\in\{1,\ldots,E\}$ is given by
\begin{eqnarray}
d_e^{\mathrm{IE}}[n] = \sqrt{\|\hat{\mathbf{t}}_e^{\mathrm E} + \Delta \mathbf{t}_e^{\mathrm E} - \mathbf{t}^{\mathrm I}[n]\|^2 + H^2}
\end{eqnarray}
and the distance between the jammer UAV and eavesdropper $e$ at time slot $n$ is given by
\begin{eqnarray}
d_e^{\mathrm{JE}}[n] = \sqrt{\|\hat{\mathbf{t}}_e^{\mathrm E} + \Delta \mathbf{t}_e^{\mathrm E} - \mathbf{t}^{\mathrm J}[n]\|^2 + H^2},
\end{eqnarray}
where $\hat{\mathbf{t}}_e^{\mathrm E} = [\hat{x}_e^{\mathrm E},\hat{y}_e^{\mathrm E}]^{\mathrm{T}} \in \mathbb{R}^{2 \times 1} $ represents the estimated location of potential eavesdropper $e$ and $\Delta \mathbf{t}_e^{\mathrm E} = [\Delta x_e^{\mathrm E}, \Delta y_e^{\mathrm E}]^{\mathrm{T}} \in \mathbb{R}^{2 \times 1} $ denotes the estimation error of $\hat{\mathbf{t}}_e^{\mathrm E}$.
Without loss of generality, we assume that the estimation error satisfies \cite{cui2018robust} 
\begin{eqnarray} \label{eqn:Delta_condition}
\| \Delta \mathbf{t}_e^{\mathrm E} \|^2 \leq (Q_e^{\mathrm E})^2,
\end{eqnarray}
where $Q_e^{\mathrm{E}}$ is the radius defining the circular uncertain region centered at the estimated location of eavesdropper $e$.
In this paper, we adopt this worst case model instead of the probabilistic model \cite{ng_EE} as the probabilistic model can be easily converted to the deterministic model under some mild conditions \cite{boshkovska2017robust}.

\subsection{UAV Power Consumption Model}

\begin{table}[t]
\centering
\scriptsize
\caption{Notations and physical meaning of variables in power consumption model.} \label{notations} 
\begin{tabular}{ c | c }
  \hline			
  Notations & Physical meaning \\ \hline
  $\Omega$ & Blade angular velocity in radians/second \\
  $r$ & Rotor radius in meter \\
  $\rho$ & Air density in $\mathrm{kg/m^3}$ \\
  $s$ & Rotor solidity in $\mathrm{m^3}$ \\
  $A_{\mathrm r}$ & Rotor disc area in $\mathrm{m^2}$ \\
  $P_o$ & Blade profile power in hovering status in watt\\
  $P_i$ & Induced power in hovering status in watt\\
  $v_0$ & Mean rotor induced velocity in forwarding flight in m/s\\
  $d_0$ & Fuselage drag ratio \\
  \hline
\end{tabular}
\end{table}

To facilitate the design of energy-efficient resource allocation, the system power consumption is defined as follows.
The flight power consumption for the rotary-wing UAV is a function of its flight velocity.
In particular, the flight power consumption models of the information UAV and the jammer UAV are given by \cite{rotary_wing_power}:
\begin{align}
P_{\mathrm{flight}}^{\mathrm I}[n] &= P_o \bigg( 1 + \frac{3 \|\mathbf{v}^{\mathrm I}[n]\|^2 }{\Omega^2 r^2} \bigg) + \frac{P_i v_0}{\| \mathbf{v}^{\mathrm I}[n] \|} \notag \\
&+ \frac{1}{2} d_0 \rho s A_{\mathrm r} \|\mathbf{v}^{\mathrm I} [n]\|^3 \,\, \text{and} \label{eqn:P_flight_I} \\
P_{\mathrm{flight}}^{\mathrm J}[n] &= P_o \bigg( 1 + \frac{3 \|\mathbf{v}^{\mathrm J}[n]\|^2 }{\Omega^2 r^2} \bigg) + \frac{P_i v_0}{\| \mathbf{v}^{\mathrm J}[n] \|} \notag \\
&+ \frac{1}{2} d_0 \rho s A_{\mathrm r} \|\mathbf{v}^{\mathrm J} [n]\|^3 , \label{eqn:P_flight_J}
\end{align}
respectively, where $\mathbf{v}^{\mathrm I}[n] = [v^{\mathrm I}_x[n], v^{\mathrm I}_y[n]]^{\mathrm{T}} \in \mathbb{R}^{2 \times 1} $ and $\mathbf{v}^{\mathrm J}[n] = [v^{\mathrm J}_x[n], v^{\mathrm J}_y[n]]^{\mathrm{T}} \in \mathbb{R}^{2 \times 1}$.
The notations and the physical meanings of the variables in \eqref{eqn:P_flight_I} and \eqref{eqn:P_flight_J} are summarized in Table \ref{notations}.
We can observe that the flight power consumption is a convex function with respect to (w.r.t.) the flight velocity for both the information UAV and the jammer UAV.
In this work, we assume that the trajectory of the jammer UAV follows a fixed path with a fixed velocity \cite{wu2018joint}.
In fact, $\mathbf{v}^{\mathrm J}[n]$ is selected by the most energy-efficient flying velocity according to the setting in \cite{rotary_wing_power}.
Since the jammer UAV is equipped with an antenna array, the beamformed artificial noise can combat the channels of eavesdroppers deliberately for providing secure communication to legitimate users.
The total power consumption of the information UAV and the jammer UAV at time slot $n$ in Joules-per-second (J/sec) includes the communication power and the flight power which can be modeled as 
\begin{eqnarray}
P_{\mathrm{total}}^{\mathrm I}[n] &=& \underset{\text{Information signals power}} {\underbrace{\sum_{k=1}^{K} \sum_{i=1}^{N_{\mathrm{F}}} \alpha_{k,i}^{\mathrm I}[n] p_{k,i}^{\mathrm I}[n]}} \zeta^{\mathrm I} + P_{\mathrm C}^{\mathrm I} \notag \\
&+& P_{\mathrm{flight}}^{\mathrm I}[n] \,\, \text{and} \\ \label{eqn:P_total_I}
P_{\mathrm{total}}^{\mathrm J}[n] &=& \underset{\text{Jamming signals power}} {\underbrace{\sum_{i=1}^{N_{\mathrm{F}}} \text{Tr}(\mathbf{Z}_i^{\mathrm J}[n])}} \zeta^{\mathrm J} + P_{\mathrm C}^{\mathrm J} + P_{\mathrm{flight}}^{\mathrm J}[n], \label{eqn:P_total_J}
\end{eqnarray}
respectively.
The constants $\zeta^{\mathrm I},\zeta^{\mathrm J}\geq1$ denote the power inefficiency of the power amplifier at the information UAV and the jammer UAV, respectively.
Variable $p_{k,i}^{\mathrm I}[n]$ denotes the information transmit power allocation for user $k$ on subcarrier $i$ at time slot $n$.
$P_{\mathrm C}^{\mathrm I}$ and $P_{\mathrm C}^{\mathrm J}$ denote the constant circuit power consumptions of the information UAV and the jammer UAV, respectively.
The binary variable $\alpha_{k,i}^{\mathrm I}[n]=1$ represents that subcarrier $i$ is assigned to user $k$ at time slot $n$.
Otherwise, $\alpha_{k,i}^{\mathrm I}[n]=0$.

\subsection{Downlink Channel Model}

\begin{figure}[t] 
  \centering
  \includegraphics[width=3.5 in]{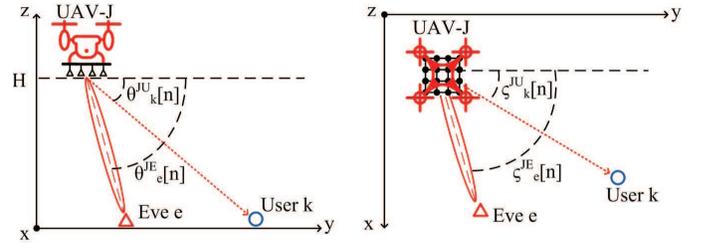} 
  \caption{Downlink LoS channel model between the jammer UAV and the ground terminals.
  The left hand side figure shows the vertical AoDs, $\theta_k^{\mathrm{JU}}[n]$ and $\theta_e^{\mathrm{JE}}[n]$, for user $k$ and eavesdropper $e$, respectively.
  The right hand side figure shows the horizontal AoDs, $\varsigma_k^{\mathrm{JU}}[n]$ and $\varsigma_e^{\mathrm{JE}}[n]$, for user $k$ and eavesdropper $e$, respectively.}
  \label{AoD}
\end{figure}

We assume that the channels from the UAVs to all ground receivers are dominated by the LoS paths and the Doppler effect is well compensated.
Thus, the channel power gain between the information UAV and user $k$ as well as eavesdropper $e$ at time slot $n$ can be characterized by the commonly adopted free-space path loss model, \cite{ DWK_2018_solar_power_comm, rotary_wing_power}, which are given by 
\begin{align}
h_k^{\mathrm{IU}}[n] &= \frac{\beta_0}{(d_k^{\mathrm{IU}}[n])^2} = \frac{\beta_0} {\|\mathbf{t}_k - \mathbf{t}^{\mathrm I}[n]\|^2 +H^2} \,\, \text{and} \\ 
h_e^{\mathrm{IE}}[n] &= \frac{\beta_0}{(d_e^{\mathrm{IE}}[n])^2} = \frac{\beta_0} {\|\hat{\mathbf{t}}_e + \Delta \mathbf{t}_e - \mathbf{t}^{\mathrm I}[n]\|^2 + H^2},
\end{align}
respectively.
The constant $\beta_0$ represents the channel power gain at a reference distance.
Besides, the channel vectors between the jammer UAV and user $k$ as well as between eavesdropper $e$ at time slot $n$ are given by equations \eqref{equation_h_k_JU} and \eqref{equation_h_e_JE} at the top of next page,  respectively\footnote{Note that $\mathbf{h}_k^{\mathrm{JU}}[n]$ and $\mathbf{h}_e^{\mathrm{JE}}[n]$ are known when the jammer UAV has a fixed trajectory.}\cite{tse2005fundamentals}, \cite{8974403}.
\begin{figure*}[t]
\begin{eqnarray}
\mathbf{h}_k^{\mathrm{JU}}[n] &=& \bigg( 1, e^{- \mathrm j \frac{2 \pi \Delta_{\mathrm J}}{\lambda_{\mathrm c}} \sin \, \theta_k^{\mathrm{JU}}[n] \cos \, \varsigma_k^{\mathrm{JU}}[n] } , \ldots e^{- \mathrm j \frac{2 \pi \Delta_{\mathrm J}}{\lambda_{\mathrm c}} \sin \, \theta_k^{\mathrm{JU}}[n] (N_{\mathrm Jx}-1) \cos \, \varsigma_k^{\mathrm{JU}}[n] } \bigg)^{\mathrm T} \notag \\
& \otimes & \bigg( 1, e^{- \mathrm j \frac{2 \pi \Delta_{\mathrm J}}{\lambda_{\mathrm c}} \sin \, \theta_k^{\mathrm{JU}}[n] \sin \, \varsigma_k^{\mathrm{JU}}[n] } , \ldots e^{- \mathrm j \frac{2 \pi \Delta_{\mathrm J}}{\lambda_{\mathrm c}} \sin \, \theta_k^{\mathrm{JU}}[n] (N_{\mathrm Jy}-1) \sin \, \varsigma_k^{\mathrm{JU}}[n] } \bigg)^{\mathrm T} \,\, \text{and} \label{equation_h_k_JU}\\
\mathbf{h}_e^{\mathrm{JE}}[n] &=& \bigg( 1, e^{- \mathrm j \frac{2 \pi \Delta_{\mathrm J}}{\lambda_{\mathrm c}} \sin \, \theta_e^{\mathrm{JE}}[n] \cos \, \varsigma_e^{\mathrm{JE}}[n] } , \ldots e^{- \mathrm j \frac{2 \pi \Delta_{\mathrm J}}{\lambda_{\mathrm c}} \sin \, \theta_e^{\mathrm{JE}}[n] (N_{\mathrm Jx}-1) \cos \, \varsigma_e^{\mathrm{JE}}[n] } \bigg)^{\mathrm T} \notag \\
& \otimes & \bigg( 1, e^{- \mathrm j \frac{2 \pi \Delta_{\mathrm J}}{\lambda_{\mathrm c}} \sin \, \theta_e^{\mathrm{JE}}[n] \sin \, \varsigma_e^{\mathrm{JE}}[n] } , \ldots e^{- \mathrm j \frac{2 \pi \Delta_{\mathrm J}}{\lambda_{\mathrm c}} \sin \, \theta_e^{\mathrm{JE}}[n] (N_{\mathrm Jy}-1) \sin \, \varsigma_e^{\mathrm{JE}}[n] } \bigg)^{\mathrm T}, \label{equation_h_e_JE}
\end{eqnarray}
\hrulefill
\end{figure*}
$\lambda_{\mathrm c}$ represents the wavelength of the carrier center frequency and $\Delta_{\mathrm J}$ is the antenna separation at the jammer UAV.
$N_{\mathrm Jx}$ and $N_{\mathrm Jy}$ represent the number of the rows and columns of the antenna array.
As shown in Figure \ref{AoD}, $\theta_k^{\mathrm{JU}}[n]$ and $\theta_e^{\mathrm{JE}}[n]$ denote the vertical angle of departure (AoD) from the jammer antenna array to user $k$ and eavesdropper $e$, respectively.
$\varsigma_k^{\mathrm{JU}}[n]$ and $\varsigma_e^{\mathrm{JE}}[n]$ denote the horizontal AoD from the jammer antenna array to user $k$ and eavesdropper $e$, respectively.
We note that
$\sin \, \theta_k^{\mathrm{JU}}[n] = \frac{H}{\sqrt{ \|\mathbf{t}_k^{\mathrm U} - \mathbf{t}^{\mathrm J}[n]\|^2 + H^2 }} $,
$\sin \, \theta_e^{\mathrm{JE}}[n] = \frac{H}{\sqrt{ \|\hat{\mathbf{t}}_e^{\mathrm E} + \Delta \mathbf{t}_e^{\mathrm E} - \mathbf{t}^{\mathrm J}[n]\|^2 + H^2 }}$,
$\sin \, \varsigma_k^{\mathrm{JU}}[n] = \frac{\|x_k^{\mathrm U}-x^{\mathrm J}[n]\|}{\|\mathbf{t}_k^{\mathrm U} - \mathbf{t}^{\mathrm J}[n]\|}$,
$\sin \, \varsigma_e^{\mathrm{JE}}[n] = \frac{\|\hat{x}_e^{\mathrm E} + \Delta x_e^{\mathrm E} - x^{\mathrm J}[n]\|} {\|\hat{\mathbf{t}}_e^{\mathrm E} + \Delta \mathbf{t}_e^{\mathrm E} - \mathbf{t}^{\mathrm J}[n]\|}$
$\cos \, \varsigma_k^{\mathrm{JU}}[n] = \frac{\|y_k^{\mathrm U}-y^{\mathrm J}[n]\|}{\|\mathbf{t}_k^{\mathrm U} - \mathbf{t}^{\mathrm J}[n]\|}$, and
$\cos \, \varsigma_e^{\mathrm{JE}}[n] = \frac{\|\hat{y}_e^{\mathrm E} + \Delta y_e^{\mathrm E} - y^{\mathrm J}[n]\|} {\|\hat{\mathbf{t}}_e^{\mathrm E} + \Delta \mathbf{t}_e^{\mathrm E} - \mathbf{t}^{\mathrm J}[n]\|}$.
Specifically, the multi-antenna wireless channel between the jammer UAV and the potential eavesdroppers captures the location uncertainty in $\text{cos} \, \varsigma_e^{\mathrm{JE}}[n]$.
For notational simplicity, we define
\begin{align}
\mathbf{H}_k^{\mathrm{JU}}[n] &= \mathbf{h}_k^{\mathrm{JU}}[n] (\mathbf{h}_k^{\mathrm{JU}}[n])^{\mathrm H} \,\, \text{and} \\
\mathbf{H}_e^{\mathrm{JE}}[n] &= \mathbf{h}_e^{\mathrm{JE}}[n] (\mathbf{h}_e^{\mathrm{JE}}[n])^{\mathrm H},
\end{align}
where $\mathbf{H}_k^{\mathrm{JU}}[n] \succeq \mathbf{0}$, $\mathbf{H}_e^{\mathrm{JE}}[n] \succeq \mathbf{0}$, $\mathbf{H}_k^{\mathrm{JU}}[n] \in \mathbb{H}^{N_{\mathrm J}}$, and $\mathbf{H}_e^{\mathrm{JE}}[n] \in \mathbb{H}^{N_{\mathrm J}}$.
Subsequently, the received interference power from the jammer UAV to users and eavesdroppers can be written as $\text{Tr}(\mathbf{H}_k^{\mathrm{JU}}[n] \mathbf{Z}_i^{\mathrm J}[n])$ and $\text{Tr}(\mathbf{H}_e^{\mathrm{JE}}[n] \mathbf{Z}_i^{\mathrm J}[n])$, respectively.

\section{Resource Allocation and Trajectory Design}
\subsection{System Achievable Rate and Energy Efficiency}

The achievable data rate for user $k$ on subcarrier $i$ at time slot $n$ is given by 
\begin{eqnarray} \label{eqn:Rate_for_user_k}
R_{k,i}^{\mathrm U}[n] = W \alpha_{k,i}^{\mathrm I}[n]  \log_2 ( 1 + \Gamma_{k,i}^{\mathrm{IU}}[n] ),
\end{eqnarray}
where $\Gamma_{k,i}^{\mathrm{IU}}[n]$ denotes the received signal-to-interference-plus-noise ratio (SINR) at user $k$ on subcarrier $i$ in time slot $n$ and it is given by
\begin{eqnarray}
\Gamma_{k,i}^{\mathrm{IU}}[n] = \frac{ p_{k,i}^{\mathrm I}[n] h_k^{\mathrm{IU}}[n] } { A_k^{\mathrm U}[n] \text{Tr}(\mathbf{H}_k^{\mathrm{JU}}[n] \mathbf{Z}_i^{\mathrm J}[n]) + W N_0},
\end{eqnarray}
where $A_k^{\mathrm U}[n] = \frac{\beta_0} {\|\mathbf{t}_k^{\mathrm U} - \mathbf{t}^{\mathrm J}[n]\|^2 +H^2}$ denotes the attenuation in the LoS path between the jammer UAV to user $k$ at time slot $n$.
$W$ represents the subcarrier bandwidth and $N_0$ is the power spectral density of the additive white Gaussian noise (AWGN).
On the other hand, the information data rate leakage to potential eavesdropper $e$ on subcarrier $i$ for user $k$ at time slot $n$ is given by 
\begin{eqnarray} \label{eqn:Rate_for_eavesdropper_e}
R_{k,e,i}^{\mathrm E}[n] = W \alpha_{k,i}^{\mathrm I}[n]  \log_2 ( 1 + \Gamma_{k,e,i}^{\mathrm{IE}}[n] ), 
\end{eqnarray}
where $\Gamma_{k,e,i}^{\mathrm{IE}}[n]$ denotes the received SINR at eavesdropper $e$ on subcarrier $i$ in time slot $n$ and it is given by 
\begin{eqnarray}
\Gamma_{k,e,i}^{\mathrm{IE}}[n] = \frac{ p_{k,i}^{\mathrm I}[n] h_e^{\mathrm{IE}}[n] } { A_e^{\mathrm E}[n] \text{Tr}(\mathbf{H}_e^{\mathrm{JE}}[n] \mathbf{Z}_i^{\mathrm J}[n]) + W N_0},
\end{eqnarray}
where $A_e^{\mathrm E}[n] = \frac{\beta_0} {\|\hat{\mathbf{t}}_e^{\mathrm E} + \Delta \mathbf{t}_e^{\mathrm E} - \mathbf{t}^{\mathrm J}[n]\|^2 + H^2}$ denotes the attenuation in the LoS path between the jammer UAV and eavesdropper $e$ at time slot $n$.
Clearly, the artificial noise generated by the jammer UAV interferes the channels of both legitimate user $k$ and eavesdropper $e$.

Thus, the system energy efficiency in bits-per-Joule (bits/J) is defined as
\begin{eqnarray} \label{eqn:ee}
\mathrm{EE}(\mathcal{A},\mathcal{P},\mathcal{Z},\mathcal{T_I},\mathcal{V_I}) = \frac{\sum_{n=1}^{N} \sum_{k=1}^{K} \sum_{i=1}^{N_{\mathrm{F}}} R_{k,i}^{\mathrm U}[n]} {\sum_{n=1}^{N} P_{\mathrm{total}}^{\mathrm I}[n] + P_{\mathrm{total}}^{\mathrm J}[n] },
\end{eqnarray}
where $\mathcal{A}=\{\alpha_{k,i}^{\mathrm I}[n], \forall k,i,n\}$ is the user scheduling variable set, $\mathcal{P}=\{p_{k,i}^{\mathrm I}[n], \forall k,i,n\}$ is the transmit power\footnote{In the considered system, although the flight power consumption is larger than the communication power, optimizing both the flight power and the communication power consumption are important to improve the system energy efficiency and to guarantee communication security.} variable set,
$\mathcal{Z}=\{\mathbf{Z}_i^{\mathrm J}[n], \forall i,n\}$ is the covariance matrix set of the artificial noises, $\mathcal{T_I}=\{\mathbf{t}^{\mathrm I}[n], \forall n\}$ is the set of the information UAV's trajectory variables, and $\mathcal{V_I}=\{\mathbf{v}^{\mathrm I}[n], \forall n\}$ is the set of the information UAV's flight velocity variables.

\subsection{Optimization Problem Formulation}

The energy-efficient design of user scheduling, transmit power allocation, UAVs' trajectory, and UAV's flight velocity is formulated as the following optimization problem\footnote{Note that the solution proposed in the paper can be easily extended to the case of 3D aviation.}: 
\begin{align} \label{opt_prob:overall}
\underset {\mathcal{A},\mathcal{P},\mathcal{Z},\mathcal{T_I},\mathcal{V_I}} {\text{maximize}} &\,\, \mathrm{EE}(\mathcal{A},\mathcal{P},\mathcal{Z},\mathcal{T_I},\mathcal{V_I}) \\
\mathrm{s.t.}\,\,\mathrm{C1}: &\,\, \alpha_{k,i}^{\mathrm I}[n] \in \{0,1\}, \forall k,i,n, \notag \\
%
%
\mathrm{C2}: &\,\, \sum_{k=1}^K \alpha_{k,i}^{\mathrm I}[n] \leq 1, \forall i,n, \notag \\
%
%
\mathrm{C3a}: &\,\, p_{k,i}^{\mathrm I}[n] \geq 0, \forall k,i,n, \notag \\
\mathrm{C3b}: &\,\, \mathbf{Z}_i^{\mathrm J}[n] \succeq \bm 0, \forall i,n, \notag \\
\mathrm{C4a}: &\,\, \sum_{k=1}^K \sum_{i=1}^{N_{\mathrm{F}}} \alpha_{k,i}^{\mathrm I}[n] p_{k,i}^{\mathrm I}[n] \leq P_{\mathrm{peak}}^{\mathrm I}, \forall n, \notag \\
\mathrm{C4b}: &\,\, \sum_{i=1}^{N_{\mathrm{F}}} \text{Tr}(\mathbf{Z}_i^{\mathrm J}[n]) \leq P_{\mathrm{peak}}^{\mathrm J}, \forall n, \notag \\
\mathrm{C5a}: &\,\, P_{\mathrm{total}}^{\mathrm I}[n] \leq  P_{\max}^{\mathrm I}, \forall n, \notag \\
\mathrm{C5b}: &\,\, P_{\mathrm{total}}^{\mathrm J}[n] \leq  P_{\max}^{\mathrm J}, \forall n,\notag \\
\mathrm{C6}: &\,\, \frac{1}{N} \sum_{n=1}^N \sum_{i=1}^{N_{\mathrm{F}}} R_{k,i}^{\mathrm U}[n] \geq R_{\min}, \forall k, \notag \\
\mathrm{C7}: &\,\, \underset{\| \Delta \mathbf{t}_e^{\mathrm{E}}\| \leq Q_e^{\mathrm{E}}}{\max} \,\, \Gamma_{k,e,i}^{\mathrm{IE}}[n] \leq \Gamma_{\mathrm{th}}, \forall k,e,i,n, \notag \\
\mathrm{C8}: &\,\, \mathbf{t}^{\mathrm I}[0] = \mathbf{t}^{\mathrm I}_0, \,\,\,\,
\mathrm{C9}: \, \mathbf{t}^{\mathrm I}[N] = \mathbf{t}^{\mathrm I}_\mathrm{F}, \notag \\
\mathrm{C10}: &\,\, \mathbf{t}^{\mathrm I}[n+1] = \mathbf{t}^{\mathrm I}[n] + \mathbf{v}^{\mathrm I}[n] \tau, n=1,...,N-1, \notag \\
\mathrm{C11}: &\,\, \|\mathbf{v}^{\mathrm I}[n]\| \leq V_{\max}^{\mathrm I}, \forall n, \notag \\
\mathrm{C12}: &\,\, \|\mathbf{v}^{\mathrm I}[n+1]-\mathbf{v}^{\mathrm I}[n]\| \leq V_{\mathrm{acc}}^{\mathrm I}, n=1,...,N-1, \notag \\
\mathrm{C13}: &\,\, \| \mathbf{t}^{\mathrm I}[n] - \mathbf{t}^{\mathrm J}[n] \|^2 \geq d_{\min}^2, \forall n. \notag 
\end{align}
Note that $\mathrm{C1}$ and $\mathrm{C2}$ are user scheduling constraints such that each subcarrier can be assigned to at most one user at each time slot to avoid multiple access interference.
$\mathrm{C3a}$ and $\mathrm{C3b}$ are the non-negative transmit power constraints for information and jammer UAVs, respectively.
$P_{\mathrm{peak}}^{\mathrm I}$ and $P_{\mathrm{peak}}^{\mathrm J}$ in $\mathrm{C4a}$ and $\mathrm{C4b}$ are the peak transmit power for the information UAV and the jammer UAV at each time slot, respectively, which is limited by the output range of the corresponding power amplifier.
Constants $P_{\max}^{\mathrm I}$ and $P_{\max}^{\mathrm J}$ in $\mathrm{C5a}$ and $\mathrm{C5b}$ are the maximum budget for total power consumption of information UAV and jammer UAV at each time slot, respectively, which are limited by the corresponding battery maximum output power.
$R_{\min}$ in $\mathrm{C6}$ denotes the minimum required individual user data rate over the whole flight duration.
$\Gamma_{\mathrm{th}}$ in $\mathrm{C7}$ is the maximum tolerable SINR threshold for eavesdropper $e$ attempting to eavesdrop the information of user $k$ on subcarrier $i$ at timeslot $n$.
Note that constraint $\mathrm{C7}$ takes into account the location uncertainty of the potential eavesdroppers.
$\mathrm{C8}$ and $\mathrm{C9}$ indicate the required UAV's initial and final locations, respectively.
$\mathrm{C10}$ draws the connections between the UAV's velocity and the displacement between two consecutive time slots for the information UAV\footnote{Note that the flight velocity of a UAV can be expressed as a function of its trajectory for a given constant time slot duration $\tau$.
Yet, expressing the flight power consumption as a function of trajectory would complicate the resource allocation design.
Therefore, we introduce the flight velocity variable $\mathbf{v}^{\mathrm I}[n]$ to simplify the problem formulation.}.
$V_{\max}^{\mathrm I}$ in $\mathrm{C11}$ is the maximum flight velocity constraint for the information UAV.
$V_{\mathrm{acc}}^{\mathrm I}$ in constraint $\mathrm{C12}$ is the maximum allowable acceleration for the information UAV in a given time slot.
$\mathrm{C13}$ limits the minimum distance between the information UAV and the jammer UAV to avoid possible collision.

\begin{Remark}
In the considered system,  secure communication can be guaranteed when $R_{\min}>\log_2(1+\Gamma_{\mathrm{th}}), \forall k,$ holds with a minimum secure rate given by $R_{\min}-\log_2(1+\Gamma_{\mathrm{th}})$.
Compared to some works directly optimizing the system secrecy rate, the parameters $R_{\min}$ and $\Gamma_{\mathrm{th}}$ in our work are chosen by the system operator which can be adopted for different applications requiring different levels of communication security.
This formulation provides flexibility in designing resource allocation algorithms and has been widely adopted, e.g. \cite{JR:Sesure_SWIPT,sun2016multi}.
\end{Remark}

\section{Problem Solution}

\begin{figure}[t]
  \centering
  \includegraphics[width=3.25 in]{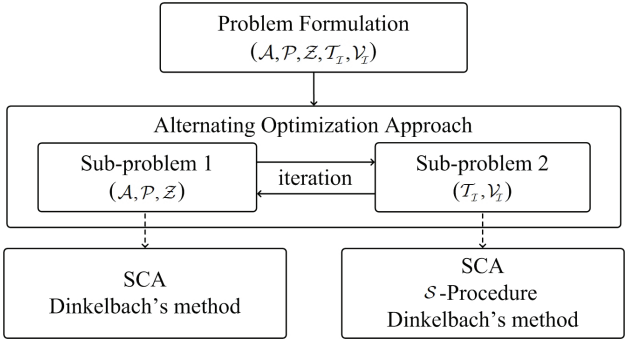} 
  \caption{A flow chart of the proposed iterative algorithm.}
  \label{algorithm_flow_chart}
\end{figure}

The formulated problem in \eqref{opt_prob:overall} is non-convex.
In general, a brute force approach may be required to obtain a globally optimal solution which is computationally intractable even for a moderate size of system.
To facilitate a low computational complexity design of resource allocation and trajectory, we aim at designing an efficient suboptimal solution.
In particular, we divide the problem \eqref{opt_prob:overall} into two sub-problems and solve them iteratively utilizing the alternating optimization to achieve a suboptimal solution of the original problem \cite{Alternating}.
Specifically, as shown in Figure \ref{algorithm_flow_chart}, sub-problem 1 optimizes the user scheduling, $\mathcal{A}$, the information transmit power allocation, $\mathcal{P}$, and the artificial noise, $\mathcal{Z}$, for a given feasible information UAV's trajectory, $\mathcal{T_I}$, and its flight velocity, $\mathcal{V_I}$.
Sub-problem 2 aims to optimize the information UAV's trajectory, $\mathcal{T_I}$, and its flight velocity, $\mathcal{V_I}$, under a given feasible user scheduling, $\mathcal{A}$, transmit power allocation, $\mathcal{P}$, and artificial noise, $\mathcal{Z}$.
The proof details on the convergence of the alternating optimization approach can be found in \cite{Alternating}.
Now, we first study the solution of sub-problem 1.

\subsection{Sub-problem 1: Optimizing User Scheduling, Communication Transmit Power Allocation, and Artificial Noise}

For a given information UAV's trajectory $\mathcal{T_I}$ and its flight velocity $\mathcal{V_I}$, we can express sub-problem 1 as the following optimization problem:
\begin{align} \label{opt_prob:sub1_origin}
\underset {\mathcal{A},\mathcal{P},\mathcal{Z}} {\text{maximize}} &\,\, \frac{\sum_{n=1}^{N} \sum_{k=1}^{K} \sum_{i=1}^{N_{\mathrm{F}}} R_{k,i}^{\mathrm U}[n]} {\sum_{n=1}^{N} \big( P_{\mathrm{total}}^{\mathrm I}[n] + P_{\mathrm{total}}^{\mathrm J}[n] \big)} \\
\mathrm{s.t.}\,\, &\,\, \mathrm{C1},\mathrm{C2},\mathrm{C3a}-\mathrm{C5a},\mathrm{C3b}-\mathrm{C5b},\mathrm{C6},\mathrm{C7}, \notag 
\end{align}
where $R_{k,i}^{\mathrm U}[n]$ in \eqref{eqn:Rate_for_user_k} is a non-convex function w.r.t. the joint optimization of $\alpha_{k,i}^{\mathrm I}[n]$, $ p_{k,i}^{\mathrm I}[n]$, and $\mathbf{Z}_i^{\mathrm J}[n]$ since the division between $ p_{k,i}^{\mathrm I}[n]$ and $\mathbf{Z}_i^{\mathrm J}[n]$.
Thus, we can rewrite it to a substraction function as 
\begin{align}
R_{k,i}^{\mathrm U}[n] &= W \alpha_{k,i}^{\mathrm I}[n] \log_2 \big( \text{Tr}(\mathbf{H}_k^{\mathrm{JU}}[n] \mathbf{Z}_i^{\mathrm J}[n]) + W N_0 \notag \\
&+ p_{k,i}^{\mathrm I}[n] h_k^{\mathrm{IU}}[n] \big)
- W \alpha_{k,i}^{\mathrm I}[n] \log_2 \big( W N_0 \notag \\
&+ \text{Tr}(\mathbf{H}_k^{\mathrm{JU}}[n] \mathbf{Z}_i^{\mathrm J}[n]) \big).
\end{align}
The problem in \eqref{opt_prob:sub1_origin} is non-convex and the non-convexity arises from the objective function and constraints $\mathrm{C1}$, $\mathrm{C4a}$, $\mathrm{C5a}$, $\mathrm{C6}$, and $\mathrm{C7}$.
In order to solve sub-problem 1 in \eqref{opt_prob:sub1_origin}, we first handle the coupling of $\alpha_{k,i}^{\mathrm I}[n] p_{k,i}^{\mathrm I}[n]$ and $\alpha_{k,i}^{\mathrm I}[n] \mathbf{Z}_i^{\mathrm J}[n]$ by introduce two auxiliary variables $\tilde{p}_{k,i}^{\mathrm I}[n] = \alpha_{k,i}^{\mathrm I}[n] p_{k,i}^{\mathrm I}[n], \forall k,i,n,$ and $\tilde{\mathbf{Z}}_{k,i}^{\mathrm J}[n] = \alpha_{k,i}^{\mathrm I}[n] \mathbf{Z}_i^{\mathrm J}[n], \forall k,i,n$.
{Then, by applying the big-M reformulation \cite{sun2018robust,lee2011mixed,bedekar2009optimum}, the couplings are resolved by introducing the following equivalent constraints:} 
\begin{align}
\mathrm{C14}: &\,\, \tilde{p}_{k,i}^{\mathrm I}[n] \leq p_{k,i}^{\mathrm I}[n], \forall k,i,n, \notag \\
\mathrm{C15}: &\,\, \tilde{p}_{k,i}^{\mathrm I}[n] \geq p_{k,i}^{\mathrm I}[n] - ( 1 - \alpha_{k,i}^{\mathrm I}[n] ) P_{\mathrm{peak}}^{\mathrm I}, \forall k,i,n, \notag \\
\mathrm{C16}: &\,\, \tilde{p}_{k,i}^{\mathrm I}[n] \geq 0, \forall k,i,n, \notag \\
\mathrm{C17}: &\,\, \tilde{p}_{k,i}^{\mathrm I}[n] \leq \alpha_{k,i}^{\mathrm I}[n] P_{\mathrm{peak}}^{\mathrm I}, \forall k,i,n, \notag \\
\mathrm{C18}: &\,\, \tilde{\mathbf{Z}}_{k,i}^{\mathrm J}[n] \preceq \mathbf{Z}_i^{\mathrm J}[n], \forall k,i,n, \notag \\
\mathrm{C19}: &\,\, \tilde{\mathbf{Z}}_{k,i}^{\mathrm J}[n] \succeq \mathbf{Z}_i^{\mathrm J}[n] - ( 1 - \alpha_{k,i}^{\mathrm I}[n] ) P_{\mathrm{peak}}^{\mathrm J} \mathbf{I}_{N_{\mathrm J}}, \forall k,i,n, \notag \\
\mathrm{C20}: &\,\, \tilde{\mathbf{Z}}_{k,i}^{\mathrm J}[n] \succeq \mathbf{0}, \forall k,i,n, \notag \\
\mathrm{C21}: &\,\, \tilde{\mathbf{Z}}_{k,i}^{\mathrm J}[n] \preceq \alpha_{k,i}^{\mathrm I}[n] P_{\mathrm{peak}}^{\mathrm J}, \forall k,i,n.
\end{align}
Then, we handle the binary user scheduling constraint $\mathrm{C1}$ in optimization problem \eqref{opt_prob:sub1_origin} by rewriting constraint $\mathrm{C1}$ in its equivalent form as: 
\begin{align}
\mathrm{C1a}: &\,\, \sum_{n=1}^N \sum_{k=1}^K \sum_{i=1}^{N_{\mathrm F}} \alpha_{k,i}^{\mathrm I}[n] - (\alpha_{k,i}^{\mathrm I}[n])^2 \leq 0, \\
\mathrm{C1b}: &\,\, 0 \leq \alpha_{k,i}^{\mathrm I}[n] \leq 1, \forall k,i,n, 
\end{align}
where $\alpha_{k,i}^{\mathrm I}[n]$ is a continuous variable with a value between zero and one.
Specifically, the continuous version of $\alpha_{k,i}^{\mathrm I} [n]$ serves as a time-sharing factor for  user $k$ in utilizing subcarrier $i$  at time slot $n$.
However, constraint $\mathrm{C1a}$ is a reverse convex function \cite{ng2016power,dinh2010local}.
In order to handle this non-convexity \cite{sun2018robust}, based on \cite{ng2016power,wei2017optimal,sun2017optimal} and for $\chi \gg 1$, the problem in \eqref{opt_prob:sub1_origin} can be equivalently transformed as:
\begin{align} \label{opt_prob:sub1_tilde}
\underset {\mathcal{A},\mathcal{P},\tilde{\mathcal{P}},\mathcal{Z},\tilde{\mathcal{Z}}} {\text{maximize}} &\,\, \frac{\sum\limits_{n=1}^{N} \sum\limits_{k=1}^{K} \sum\limits_{i=1}^{N_{\mathrm{F}}} \tilde{R}_{k,i}^{\mathrm U} [n] - \chi\big(\alpha_{k,i}^{\mathrm I}[n] - (\alpha_{k,i}^{\mathrm I}[n])^2\big)} { \sum_{n=1}^{N} ( \tilde{P}_{\mathrm{total}}^{\mathrm I}[n] + P_{\mathrm{total}}^{\mathrm J}[n])} \\
\mathrm{s.t.} &\,\, \mathrm{C1b}, \mathrm{C2}, \mathrm{C3a}, \mathrm{C3b}-\mathrm{C5b}, \mathrm{C7}, \mathrm{C14}-\mathrm{C21}, \notag \\ 
\widetilde{\mathrm{C4}}\mathrm a : &\,\, \sum_{k=1}^K \sum_{i=1}^{N_{\mathrm F}} \tilde{p}_{k,i}^{\mathrm I}[n] \leq P_{\mathrm{peak}}^{\mathrm I}, \forall n, \notag \\
\widetilde{\mathrm{C5}}\mathrm a : &\,\, \tilde{P}_{\mathrm{total}}^{\mathrm I}[n] \leq P_{\max}^{\mathrm I}, \forall n, \notag \\ 
\widetilde{\mathrm{C6}} : &\,\, \frac{1}{N} \sum_{n=1}^N \sum_{i=1}^{N_{\mathrm{F}}} \tilde{R}_{k,i}^{\mathrm U}[n] \geq R_{\min}, \forall k, \notag 
\end{align}
where $\tilde{\mathcal{P}}=\{\tilde{p}_{k,i}^{\mathrm I}[n], \forall k,i,n\}$, $\tilde{\mathcal{Z}}=\{\tilde{\mathbf{Z}}_{k,i}^{\mathrm J}[n], \forall k,i,n\}$, 
\begin{align}
\tilde{R}_{k,i}^{\mathrm U}[n] &= D^{\mathrm{I}}_{k,i}[n] - D^{\mathrm{II}}_{k,i}[n] \\ 
D^{\mathrm{I}}_{k,i}[n] &= W \alpha_{k,i}^{\mathrm I}[n] \log_2 \bigg( W N_0 \notag \\
&+\frac{\text{Tr}(\mathbf{H}_k^{\mathrm{JU}}[n] \tilde{\mathbf{Z}}_{k,i}^{\mathrm J}[n]) + \tilde{p}_{k,i}^{\mathrm I}[n] h_k^{\mathrm{IU}}[n]}{\alpha_{k,i}^{\mathrm I}[n]} \bigg), \forall k,i,n, \\
D^{\mathrm{II}}_{k,i}[n] &= W \alpha_{k,i}^{\mathrm I}[n] \log_2 \bigg( \frac{ \text{Tr}(\mathbf{H}_k^{\mathrm{JU}}[n] \tilde{\mathbf{Z}}_{k,i}^{\mathrm J}[n])} {\alpha_{k,i}^{\mathrm I}[n]} \notag \\
&+ W N_0 \bigg), \forall k,i,n, \,\, \text{and} \\
\tilde{P}_{\mathrm{total}}^{\mathrm I}[n] &= \sum_{k=1}^{K} \sum_{i=1}^{N_{\mathrm{F}}} \tilde{p}_{k,i}^{\mathrm I}[n] + P_{\mathrm C}^{\mathrm I} + P_{\mathrm{flight}}^{\mathrm I}[n].
\end{align}
The variable $\chi \gg 1$ acts as a penalty factor for accounting the objective function for any $\alpha_{k,i}^{\mathrm I}[n]$ that is not equal to 0 or 1.
Note that the problem in \eqref{opt_prob:sub1_tilde} is still non-convex and the non-convexity arises from the objective function and constraint $\widetilde{\mathrm{C6}}$.
Thus, we handle the data rate in the objective function and constraint $\widetilde{\mathrm{C6}}$ since it is the difference of convex (DC) functions.
Based on the SCA and \cite{wu2018joint,wei2017optimal}, for given feasible points $(\alpha_{k,i}^{\mathrm I}[n])^{j^{\mathrm{A1}}}$ and $(\tilde{\mathbf{Z}}_{k,i}^{\mathrm J}[n])^{j^{\mathrm{A1}}}$, a lower bound of the data rate can be obtained by its first-order Taylor expansion as 
\begin{align} \label{eqn:R_tilde-lb_j_Algo1}
\tilde{R}_{k,i}^{\mathrm U}[n] &\geq (\tilde{R}_{k,i,\mathrm{lb}}^{\mathrm U}[n])^{j^{\mathrm{A1}}} = D^{\mathrm{I}}_{k,i}[n] - (D^{\mathrm{II}}_{k,i,\mathrm{ub}}[n])^{j^{\mathrm{A1}}} \notag \\
&= D^{\mathrm{I}}_{k,i}[n] - \big[(D^{\mathrm{II}}_{k,i}[n])^{j^{\mathrm{A1}}} + \nabla_{\mathcal{A}} D^{\mathrm{II}}_{k,i}[n] \notag \\
&\times (\alpha_{k,i}^{\mathrm I}[n] - (\alpha_{k,i}^{\mathrm I}[n])^{j^{\mathrm{A1}}}) + \nabla_{\{\tilde{\mathcal{Z}}\}_{r,c}} D^{\mathrm{II}}_{k,i}[n] \notag \\
&\times (\{\tilde{\mathbf{Z}}_{k,i}^{\mathrm J}[n]\}_{r,c} - \{(\tilde{\mathbf{Z}}_{k,i}^{\mathrm J}[n])^{j^{\mathrm{A1}}}\}_{r,c}) \big],
\end{align}
where $r \in \{1,...,N_{\mathrm Jx} N_{\mathrm Jy}\}$, $c \in \{1,...,N_{\mathrm Jx} N_{\mathrm Jy}\}$, $(D^{\mathrm{II}}_{k,i,\mathrm{ub}}[n])^{j^{\mathrm{A1}}} $, $\nabla_{\mathcal{A}} D^{\mathrm{II}}_{k,i}[n] (\alpha_{k,i}^{\mathrm I}[n] - (\alpha_{k,i}^{\mathrm I}[n])^{j^{\mathrm{A1}}})$, and $\nabla_{\{\tilde{\mathcal{Z}}\}_{r,c}} D^{\mathrm{II}}_{k,i}[n] (\{\tilde{\mathbf{Z}}_{k,i}^{\mathrm J}[n]\}_{r,c} - \{(\tilde{\mathbf{Z}}_{k,i}^{\mathrm J}[n])^{j^{\mathrm{A1}}}\}_{r,c})$ are given by equations \eqref{equation_DC_D2j}, \eqref{equation_DC_nabla_a}, and \eqref{equation_DC_nabla_z} at the top of next page, respectively.
\begin{figure*}
\begin{eqnarray}
&& (D^{\mathrm{II}}_{k,i}[n])^{j^{\mathrm{A1}}} = W (\alpha_{k,i}^{\mathrm I}[n])^{j^{\mathrm{A1}}} \log_2 \bigg( \frac{ \text{Tr}(\mathbf{H}_k^{\mathrm{JU}}[n] (\tilde{\mathbf{Z}}_{k,i}^{\mathrm J}[n])^{j^{\mathrm{A1}}})} {(\alpha_{k,i}^{\mathrm I}[n])^{j^{\mathrm{A1}}}} + W N_0 \bigg), \forall k,i,n, \label{equation_DC_D2j} \\
&& \nabla_{\mathcal{A}} D^{\mathrm{II}}_{k,i}[n] (\alpha_{k,i}^{\mathrm I}[n] - (\alpha_{k,i}^{\mathrm I}[n])^{j^{\mathrm{A1}}}) = W \log_2 \bigg( \frac{ \text{Tr}(\mathbf{H}_k^{\mathrm{JU}}[n] (\tilde{\mathbf{Z}}_{k,i}^{\mathrm J}[n])^{j^{\mathrm{A1}}})} {(\alpha_{k,i}^{\mathrm I}[n])^{j^{\mathrm{A1}}}} + W N_0 \bigg) (\alpha_{k,i}^{\mathrm I}[n] - (\alpha_{k,i}^{\mathrm I}[n])^{j^{\mathrm{A1}}}) \notag \\
&& - \frac{W \text{Tr}\big(\mathbf{H}_k^{\mathrm{JU}}[n] (\tilde{\mathbf{Z}}_{k,i}^{\mathrm J}[n])^{j^{\mathrm{A1}}} \big) (\alpha_{k,i}^{\mathrm I}[n] - (\alpha_{k,i}^{\mathrm I}[n])^{j^{\mathrm{A1}}}) }{(\text{Tr}(\mathbf{H}_k^{\mathrm{JU}}[n] (\tilde{\mathbf{Z}}_{k,i}^{\mathrm J}[n])^{j^{\mathrm{A1}}}) + W N_0 (\alpha_{k,i}^{\mathrm I}[n])^{j^{\mathrm{A1}}}) \ln 2}, \forall k,i,n, \,\, \text{and} \label{equation_DC_nabla_a} \\
&& \nabla_{\{\tilde{\mathcal{Z}}\}_{r,c}} D^{\mathrm{II}}_{k,i}[n] (\{\tilde{\mathbf{Z}}_{k,i}^{\mathrm J}[n]\}_{r,c} - \{(\tilde{\mathbf{Z}}_{k,i}^{\mathrm J}[n])^{j^{\mathrm{A1}}}\}_{r,c}) \notag \\
&& = \frac{ W (\alpha_{k,i}^{\mathrm I}[n])^{j^{\mathrm{A1}}} \{\mathbf{H}_{k}^{\mathrm{JU}}[n]\}_{c,r} (\{\tilde{\mathbf{Z}}_{k,i}^{\mathrm J}[n]\}_{r,c} - \{(\tilde{\mathbf{Z}}_{k,i}^{\mathrm J}[n])^{j^{\mathrm{A1}}}\}_{r,c})}{(\text{Tr}(\mathbf{H}_k^{\mathrm{JU}}[n] (\tilde{\mathbf{Z}}_{k,i}^{\mathrm J}[n])^{j^{\mathrm{A1}}}) + W N_0 (\alpha_{k,i}^{\mathrm I}[n])^{j^{\mathrm{A1}}}) \ln 2}, \forall k,i,n,r,c. \label{equation_DC_nabla_z}
\end{eqnarray}
\hrulefill
\end{figure*}
Similarly, we can obtain an upper bound of the penalty part as 
\begin{align} \label{eqn:A_ub_j_Algo1}
& \alpha_{k,i}^{\mathrm I}[n] - (\alpha_{k,i}^{\mathrm I}[n])^2 \leq (A_{k,i,\mathrm{ub}}[n])^{j^{\mathrm{A1}}} \notag\\
= & \alpha_{k,i}^{\mathrm I}[n] - \big( (\alpha_{k,i}^{\mathrm I}[n])^{j^{\mathrm{A1}}} \big)^2 \notag \\
+ & 2 (\alpha_{k,i}^{\mathrm I}[n])^{j^{\mathrm{A1}}} \big( \alpha_{k,i}^{\mathrm I}[n] - (\alpha_{k,i}^{\mathrm I}[n])^{j^{\mathrm{A1}}} \big).
\end{align}
Then, we handle constraint $\mathrm{C7}$ in \eqref{opt_prob:sub1_tilde} by considering its subset:
\begin{eqnarray} \label{eqn:max_max/min}
&&\underset{\| \Delta \mathbf{t}_e^{\mathrm{E}}\| \leq Q_e^{\mathrm{E}}}{\max} \,\, \frac{ p_{k,i}^{\mathrm I}[n] h_e^{\mathrm{IE}}[n] } { \text{Tr}(\mathbf{H}_e^{\mathrm{JE}}[n] \mathbf{Z}_i^{\mathrm J}[n]) + W N_0} \notag \\
&\leq& \underset{\| \Delta \mathbf{t}_e^{\mathrm{E}}\| \leq Q_e^{\mathrm{E}}}{\max} \,\, \frac{ p_{k,i}^{\mathrm I}[n] h_e^{\mathrm{IE}}[n] } { \underset{\| \Delta \mathbf{t}_e^{\mathrm{E}}\| \leq Q_e^{\mathrm{E}}}{\min} \,\, \text{Tr}(\mathbf{H}_e^{\mathrm{JE}}[n] \mathbf{Z}_i^{\mathrm J}[n]) + W N_0} \notag \\ 
&=& \frac{\underset{\| \Delta \mathbf{t}_e^{\mathrm{E}}\| \leq Q_e^{\mathrm{E}}}{\max} \,\, p_{k,i}^{\mathrm I}[n] h_e^{\mathrm{IE}}[n] } { \underset{\| \Delta \mathbf{t}_e^{\mathrm{E}}\| \leq Q_e^{\mathrm{E}}}{\min} \,\, \text{Tr}(\mathbf{H}_e^{\mathrm{JE}}[n] \mathbf{Z}_i^{\mathrm J}[n]) + W N_0} \leq \Gamma_{\mathrm{th}}.
\end{eqnarray}
{This safe approximation \cite{ben2009robust,le2017robust} imposes a more stringent constraint on the leakage SINR and solving the corresponding problem provides a performance lower bound of the original problem.}

After applying \eqref{eqn:R_tilde-lb_j_Algo1}-\eqref{eqn:max_max/min} to \eqref{opt_prob:sub1_tilde}, a suboptimal solution of \eqref{opt_prob:sub1_tilde} can be obtained by solving
\begin{align} \label{opt_prob:sub1_lb}
\underset {\mathcal{A},\mathcal{P},\tilde{\mathcal{P}},\mathcal{Z},\tilde{\mathcal{Z}}} {\text{maximize}} &\,\, \frac{\sum\limits_{n=1}^{N} \sum\limits_{k=1}^{K} \sum\limits_{i=1}^{N_{\mathrm{F}}} (\tilde{R}_{k,i,\mathrm{lb}}^{\mathrm U}[n])^{j^{\mathrm{A1}}} - \chi (A_{k,i,\mathrm{ub}}[n])^{j^{\mathrm{A1}}}} {\sum_{n=1}^{N} ( \tilde{P}_{\mathrm{total}}^{\mathrm I}[n] + P_{\mathrm{total}}^{\mathrm J}[n])} \\
 \mathrm{s.t.} &\,\, \mathrm{C1b}, \mathrm{C2}, \mathrm{C3a}, \widetilde{\mathrm{C4}}\mathrm a, \widetilde{\mathrm{C5}}\mathrm a, \mathrm{C3b}-\mathrm{C5b}, \mathrm{C14}-\mathrm{C21}, \notag\\
\widetilde{\widetilde{\mathrm{C6}}}:&\,\, \frac{1}{N} \sum_{n=1}^N  \sum_{i=1}^{N_{\mathrm{F}}} (\tilde{R}_{k,i,\mathrm{lb}}^{\mathrm U}[n])^{j^{\mathrm{A1}}} \geq R_{\min}, \forall k, \notag \\
\widetilde{\mathrm{C7}}:&\,\, p_{k,i}^{\mathrm I}[n] \underset{\| \Delta \mathbf{t}_e^{\mathrm{E}}\| \leq Q_e^{\mathrm{E}}}{\max} h_e^{\mathrm{IE}}[n] \notag \\
&\,\, \leq \Gamma_{\mathrm{th}} \big( \text{Tr}(\underset{ \| \Delta \mathbf{t}_e^{\mathrm{E}}\| \leq Q_e^{\mathrm{E}} }{\min} \mathbf{H}_e^{\mathrm{JE}}[n] \mathbf{Z}_i^{\mathrm J}[n]) \notag \\
&\,\, + W N_0 \big), \forall k,e,i,n. \notag
\end{align}
Then, for improving the quality of the obtained suboptimal solutions, we update the feasible solution, $(\alpha_{k,i}^{\mathrm I}[n])^{j^{\mathrm{A1}}}$ and $(\tilde{\mathbf{Z}}_{k,i}^{\mathrm J}[n])^{j^{\mathrm{A1}}}$, obtained by solving \eqref{opt_prob:sub1_lb} in the SCA iteratively, cf. Main loop in $\textbf{Algorithm \ref{alg_sub_1}}$.

\begin{table}[t] 
\scriptsize
\linespread{1}
\begin{algorithm} [H]
\caption{Proposed Algorithm for Solving Sub-problem 1} \label{alg_sub_1}
\begin{algorithmic} [1]

\STATE Initialize the convergence tolerance $\epsilon_1 \rightarrow 0$, the maximum number of iterations for main loop $J^{\mathrm{A1}}_{\max}$, the initial iteration index $j^{\mathrm{A1}} = 1$, and the initial system energy efficiency $q_1^{j^{\mathrm{A1}}} = 0$

\REPEAT[Main Loop: SCA]

\STATE Set $j^{\mathrm{A1}} = j^{\mathrm{A1}} +1$
\STATE Using $\textbf{Algorithm \ref{dinkelbach}}$ to obtain $\{\underline{\mathcal{A}}^{(j^{\mathrm{A1}})}$, $\underline{\mathcal{P}}^{(j^{\mathrm{A1}})}$, $\underline{\tilde{\mathcal{P}}}^{(j^{\mathrm{A1}})}$, $\underline{\mathcal{Z}}^{(j^{\mathrm{A1}})}$, $\underline{\tilde{\mathcal{Z}}}^{(j^{\mathrm{A1}})}\}$ and $q_1^{(j^{\mathrm{A1}})}$

\UNTIL{$j^{\mathrm{A1}} = J^{\mathrm{A1}}_{\max}$ or $\frac{|q_1^{(j^{\mathrm{A1}})} - q_1^{(j^{\mathrm{A1}}+1)}|}{q_1^{(j^{\mathrm{A1}})}} \leq \epsilon_1$}
\STATE Return $\{\mathcal{A}^*$, $\mathcal{P}^*$, $\tilde{\mathcal{P}}^*$, $\mathcal{Z}^*$, $\tilde{\mathcal{Z}}^*\}$ = $\{\underline{\mathcal{A}}^{(j^{\mathrm{A1}})}$, $\underline{\mathcal{P}}^{(j^{\mathrm{A1}})}$, $\underline{\tilde{\mathcal{P}}}^{(j^{\mathrm{A1}})}$, $\underline{\mathcal{Z}}^{(j^{\mathrm{A1}})}$, $\underline{\tilde{\mathcal{Z}}}^{(j^{\mathrm{A1}})}\}$ and $q_1^* = q_1^{(j^{\mathrm{A1}})}$

\end{algorithmic}
\end{algorithm}
\end{table}

Now, we discuss the methodology for solving sub-problem 1 in \eqref{opt_prob:sub1_lb}.
In particular, we tackle the fractional form objective function in \eqref{opt_prob:sub1_lb}.
Let $q_1^*$ be the maximum system energy efficiency of sub-problem 1 which is given by 
\begin{eqnarray}
q_1^* = \frac{R(\mathcal{A}^*,\tilde{\mathcal{P}}^*,\tilde{\mathcal{Z}}^*)} {P(\tilde{\mathcal{P}}^*,\mathcal{Z}^*)} = \underset {\mathcal{A},\mathcal{P},\tilde{\mathcal{P}},\mathcal{Z},\tilde{\mathcal{Z}} \in \mathcal{F}} {\text{maximize}}\,\, \frac{R(\mathcal{A},\tilde{\mathcal{P}},\tilde{\mathcal{Z}})} {P(\tilde{\mathcal{P}},\mathcal{Z})},
\end{eqnarray}
where $\mathcal{A}^*$, $\mathcal{P}^*$, $\tilde{\mathcal{P}}^*$, $\mathcal{Z}^*$, and $\tilde{\mathcal{Z}}^*$ are the optimal value sets of the optimization variables in \eqref{opt_prob:sub1_tilde}.
$\mathcal{F}$ is the feasible solution set spanned by constraints $\mathrm{C1b},\mathrm{C2},\mathrm{C3a}, \widetilde{\mathrm{C4}}\mathrm a, \widetilde{\mathrm{C5}}\mathrm a,\mathrm{C3b}-\mathrm{C5b},\widetilde{\widetilde{\mathrm{C6}}}, \widetilde{\mathrm{C7}}$, and $\mathrm{C14}-\mathrm{C21}$.
Now, by applying the fractional programming theory \cite{ng_EE}, the objective function of \eqref{opt_prob:sub1_lb} can be equivalently transformed into a subtractive form.
In particular, the optimal value of $q_1^*$ in \eqref{opt_prob:sub1_lb} can be achieved if and only if 
\begin{eqnarray}
& \underset {\mathcal{A},\mathcal{P},\tilde{\mathcal{P}},\mathcal{Z},\tilde{\mathcal{Z}} \in \mathcal{F}} {\text{maximize}} R(\mathcal{A},\tilde{\mathcal{P}},\tilde{\mathcal{Z}}) - q_1^* P(\tilde{\mathcal{P}},\mathcal{Z}) \notag \\
& = R(\mathcal{A}^*,\tilde{\mathcal{P}}^*,\tilde{\mathcal{Z}}^*) - q_1^* P(\tilde{\mathcal{P}}^*,\mathcal{Z}^*) = 0,
\end{eqnarray}
for $R(\mathcal{A},\tilde{\mathcal{P}},\tilde{\mathcal{Z}}) \geq 0$ and $P(\mathcal{P},\mathcal{Z}) > 0$.

Therefore, we can apply the iterative Dinkelbach's method \cite{dinkelbach} to solve \eqref{opt_prob:sub1_lb}.
In particular, for the $j^{\mathrm{A1}}$-th iteration for sub-problem 1 and a given intermediate value $q_1^{(j^{\mathrm{A1}}_{\mathrm{in}})}$, we need to solve a convex optimization as follows: 
\begin{eqnarray} \label{opt_prob:sub1_final}
&&\{\underline{\mathcal{A}},\underline{\mathcal{P}},\underline{\tilde{\mathcal{P}}},\underline{\mathcal{Z}},\underline{\tilde{\mathcal{Z}}}\} \\
&&= \arg \, \underset {\mathcal{A},\mathcal{P},\tilde{\mathcal{P}},\mathcal{Z},\tilde{\mathcal{Z}}} {\text{maximize}} \,\, \sum_{n=1}^{N} \sum_{k=1}^{K} \sum_{i=1}^{N_{\mathrm{F}}} (\tilde{R}_{k,i,\mathrm{lb}}^{\mathrm U}[n])^{j^{\mathrm{A1}}} \notag \\
&& - \chi (A_{k,i,\mathrm{ub}}[n])^{j^{\mathrm{A1}}} - q_1^{(j^{\mathrm{A2}}_{\mathrm{in}})} \sum_{n=1}^{N} (\tilde{P}_{\mathrm{total}}^{\mathrm I}[n] + P_{\mathrm{total}}^{\mathrm J}[n]) \notag \\
&&\mathrm{s.t.}\,\, \mathrm{C1b}, \mathrm{C2}, \mathrm{C3a}-\mathrm{C5a}, \mathrm{C3b}-\mathrm{C5b}, \widetilde{\widetilde{\mathrm{C6}}}, \widetilde{\mathrm{C7}}, \notag \\
&& \,\, \,\, \,\, \,\, \,\, \mathrm{C14}-\mathrm{C21}, \notag 
\end{eqnarray}
where $\{\underline{\mathcal{A}},\underline{\mathcal{P}},\underline{\tilde{\mathcal{P}}},\underline{\mathcal{Z}},\underline{\tilde{\mathcal{Z}}}\}$ is the optimal solution of \eqref{opt_prob:sub1_final} for a given $q_1^{(j^{\mathrm{A2}}_{\mathrm{in}})}$.
Then, the intermediate energy efficiency value $q_1^{(j^{\mathrm{A2}}_{\mathrm{in}})}$ should be updated as $q_1^{(j^{\mathrm{A2}}_{\mathrm{in}})} = \frac{R(\underline{\mathcal{A}}, \underline{\tilde{\mathcal{P}}}, \underline{\tilde{\mathcal{Z}}})} {P(\underline{\tilde{\mathcal{P}}},\underline{\mathcal{Z}})}$ for each iteration of the Dinkelbach's method until convergence\footnote{Note that the convergence of the Dinkelbach's method is guaranteed if the problem in \eqref{opt_prob:sub1_final} can be solved optimally in each iteration \cite{dinkelbach}.}.
Sine the problem in \eqref{opt_prob:sub1_final} is jointly convex w.r.t. the optimization variables, it can be solved efficiently via convex programm solvers, e.g. CVX \cite{cvx}.
On the other hand, it is interesting to study structure of the generated artificial noise which is summarised in the following theorem.

\begin{Thm} \label{thm:rank_Z_leq_1}
If the optimization problem in \eqref{opt_prob:sub1_final} is feasible, the rank of the optimal artificial noise matrix $\mathrm{Rank}(\mathbf{Z}) \leq 1$.
\end{Thm}  
\emph{\quad Proof: } Please refer to the Appendix. \qed

\noindent
Although there are multiple eavesdroppers in the system, rank-one beamforming is optimal for \eqref{opt_prob:sub1_final} to guarantee secure and energy efficient communication.

\begin{table}[t] 
\scriptsize
\linespread{1}
\begin{algorithm} [H]
\caption{Dinkelbach's Method} \label{dinkelbach}
\begin{algorithmic} [1]

\STATE Initialize the convergence tolerance $\epsilon_2 \rightarrow 0$, the maximum number of iterations $J_{\mathrm{in},\max}^{\mathrm{A2}}$, the iteration index $j_{\mathrm{in}}^{\mathrm{A2}} = 1$, and the initial system energy efficiency $q_1^{(j_{\mathrm{in}}^{\mathrm{A2}})}=0$

\REPEAT[Inner Loop: Dinkelbach Method]
\STATE Solve \eqref{opt_prob:sub1_final} for the given $q_1^{(j_{\mathrm{in}}^{\mathrm{A2}})}$ to obtain  $\{\underline{\mathcal{A}}^{(j_{\mathrm{in}}^{\mathrm{A2}})}, \underline{\mathcal{P}}^{(j_{\mathrm{in}}^{\mathrm{A2}})}, \underline{\tilde{\mathcal{P}}}^{(j_{\mathrm{in}}^{\mathrm{A2}})}, \underline{\mathcal{Z}}^{(j_{\mathrm{in}}^{\mathrm{A2}})}, \underline{\tilde{\mathcal{Z}}}^{(j_{\mathrm{in}}^{\mathrm{A2}})}\}$

\IF{$R(\underline{\mathcal{A}}^{(j_{\mathrm{in}}^{\mathrm{A2}})}$, $\underline{\tilde{\mathcal{P}}}^{(j_{\mathrm{in}}^{\mathrm{A2}})}$, $\underline{\tilde{\mathcal{Z}}}^{(j_{\mathrm{in}}^{\mathrm{A2}})})$ - $q_1^{j^{\mathrm{A2}}_{\mathrm{in}}}$ $P(\underline{\tilde{\mathcal{P}}}^{(j_{\mathrm{in}}^{\mathrm{A2}})}$, $\underline{\mathcal{Z}}^{(j_{\mathrm{in}}^{\mathrm{A2}})}) < \epsilon_2$}

\STATE  Inner Loop Convergence $=$ \TRUE

\RETURN $\{\underline{\mathcal{A}}^{(j^{\mathrm{A1}})}$, $\underline{\mathcal{P}}^{(j^{\mathrm{A1}})}$, $\underline{\tilde{\mathcal{P}}}^{(j^{\mathrm{A1}})}$, $\underline{\mathcal{Z}}^{(j^{\mathrm{A1}})}$, $\underline{\tilde{\mathcal{Z}}}^{(j^{\mathrm{A1}})}\}$ $= \{\underline{\mathcal{A}}^{(j_{\mathrm{in}}^{\mathrm{A2}})}$, $\underline{\mathcal{P}}^{(j_{\mathrm{in}}^{\mathrm{A2}})}$, $\underline{\tilde{\mathcal{P}}}^{(j_{\mathrm{in}}^{\mathrm{A2}})}$, $\underline{\mathcal{Z}}^{(j_{\mathrm{in}}^{\mathrm{A2}})}$, $\underline{\tilde{\mathcal{Z}}}^{(j_{\mathrm{in}}^{\mathrm{A2}})}\}$ and $q_1^{\mathrm{A1}} = \frac{R(\underline{\mathcal{A}}^{(j_{\mathrm{in}}^{\mathrm{A2}})}, \underline{\tilde{\mathcal{P}}}^{(j_{\mathrm{in}}^{\mathrm{A2}})}, \underline{\tilde{\mathcal{Z}}}^{(j_{\mathrm{in}}^{\mathrm{A2}})})} {P(\underline{\tilde{\mathcal{P}}}^{(j_{\mathrm{in}}^{\mathrm{A2}})}, \underline{\mathcal{Z}}^{(j_{\mathrm{in}}^{\mathrm{A2}})})}$

\ELSE
\STATE Set $q_1^{(j_{\mathrm{in}}^{\mathrm{A2}}+1)} = \frac{R(\underline{\mathcal{A}}^{(j_{\mathrm{in}}^{\mathrm{A2}})}, \underline{\tilde{\mathcal{P}}}^{(j_{\mathrm{in}}^{\mathrm{A2}})}, \underline{\tilde{\mathcal{Z}}}^{(j_{\mathrm{in}}^{\mathrm{A2}})})} {P(\underline{\tilde{\mathcal{P}}}^{(j_{\mathrm{in}}^{\mathrm{A2}})}, \underline{\mathcal{Z}}^{(j_{\mathrm{in}}^{\mathrm{A2}})})}$ and $j_{\mathrm{in}}^{\mathrm{A2}} = j_{\mathrm{in}}^{\mathrm{A2}}+1$
\STATE  Inner Loop Convergence $=$ \FALSE

\ENDIF

\UNTIL{Inner Loop Convergence $=$ \TRUE $\,$ or $j_{\mathrm{in}}^{\mathrm{A2}} = J_{\mathrm{in},\max}^{\mathrm{A2}}$}

\end{algorithmic}
\end{algorithm}
\end{table}

The proposed algorithm for solving sub-problem 1 is summarized in $\textbf{Algorithm \ref{alg_sub_1}}$ which consists of two nested loops.
Specifically, in each iteration of the main loop, we solve the inner loop problem, i.e., lines 2-11 of $\textbf{Algorithm \ref{dinkelbach}}$, in \eqref{opt_prob:sub1_final} for a given parameter $q_1^{(j^{\mathrm{A2}}_{\mathrm{in}})}$ given by the initialization or last iteration.
After obtaining the solution in the inner loop via the Dinkelbach's method, we update parameter $q_1^{(j^{\mathrm{A2}}_{\mathrm{in}})}$ and use it for solving the inner loop problem in the next iteration.
This procedure is repeated until the proposed algorithm converges.
We note that the convergence of the SCA is guaranteed \cite{EE_fixed_wing}.

\subsection{Sub-problem 2: Optimizing Information UAV's Trajectory and Flight Velocity}

For a given user scheduling $\mathcal{A} = \{ \alpha_{k,i}^{\mathrm I}[n], \forall k,i,n \}$, information transmit power allocation $\mathcal{P} = \{ p_{k,i}^{\mathrm I}[n], \forall k,i,n\}$, and jammer UAV's artificial noise $\mathcal{Z}=\{\mathbf{Z}_i^{\mathrm J}[n], \forall i,n\}$, we can express sub-problem 2 as 
\begin{align} \label{opt_prob:sub2_origin}
\underset {\mathcal{T_I},\mathcal{V_I}} {\text{maximize}} &\,\,  \frac{\sum_{n=1}^{N} \sum_{k=1}^{K} \sum_{i=1}^{N_{\mathrm{F}}} R_{k,i}^{\mathrm U}[n]} {\sum_{n=1}^{N} (P_{\mathrm{total}}^{\mathrm I}[n] + P_{\mathrm{total}}^{\mathrm J}[n] )} \\
\mathrm{s.t.}\,\, &\,\, \mathrm{C5a},\mathrm{C6},\mathrm{C7},\mathrm{C8} - \mathrm{C13}. \notag
\end{align}
The problem in \eqref{opt_prob:sub2_origin} is non-convex and the non-convexity arises from the objective function and constraints $\mathrm{C6}$ and $\mathrm{C7}$.
To facilitate the solution design, we introduce two slack optimization variables $u_k[n]$ and $\upsilon^{\mathrm I}[n]$ to transform the problem into its equivalent form as follows: 
\begin{align} \label{opt_prob:sub2_bar}
\underset {\mathcal{T_I},\mathcal{V_I},\,\mathcal{U_K},{\bm\Upsilon_{\mathcal I}}} {\text{maximize}} &\,\, \frac{\sum_{n=1}^N \sum_{k=1}^K \sum_{i=1}^{N_{\mathrm{F}}} \bar{R}_{k,i}^{\mathrm U}[n] } {\sum_{n=1}^{N} (\bar{P}_{\mathrm{total}}^{\mathrm I}[n] + P_{\mathrm{total}}^{\mathrm J}[n] )} \\
\mathrm{s.t.}\,\, &\,\, \mathrm{C5a},\mathrm{C8} - \mathrm{C13}, \notag \\
\overline{\mathrm{C6}}: &\,\, \frac{1}{N} \sum_{n=1}^N \sum_{i=1}^{N_{\mathrm{F}}} \bar{R}_{k,i}^{\mathrm U}[n] \geq R_{\min}, \forall k, \notag \\
\overline{\mathrm{C7}}: &\,\, \underset{\| \Delta \mathbf{t}_e^{\mathrm{E}}\| \leq Q_e^{\mathrm{E}}}{\min} \,\, \|\mathbf{t}_e^{\mathrm E} + \Delta \mathbf{t}_e^{\mathrm E} - \mathbf{t}^{\mathrm I}[n]\|^2 + H^2 \notag \\
&\,\, \geq \frac{\gamma_{k,e,i}^{\mathrm{IJE}}[n]} {\Gamma_{\mathrm{th}}}, \forall k,e,i,n, \notag \\
\mathrm{C22}: &\,\, \|\mathbf{t}_k^{\mathrm U}-\mathbf{t}^{\mathrm I}[n]\|^2 + H^2 \leq u_k[n], \forall k,n, \notag \\
\mathrm{C23}: &\,\, \|\mathbf{v}^{\mathrm I}[n]\| ^2 \geq (\upsilon^{\mathrm I}[n])^2, \forall n, \notag \\
\mathrm{C24}: &\,\, \upsilon^{\mathrm I}[n] \geq 0, \forall n, \notag 
\end{align}
where $\mathcal{U_K}=\{u_k[n], \forall k,n\}$, ${\bm\Upsilon_{\mathcal I}} = \{\upsilon^{\mathrm I}[n], \forall n\}$,
\begin{align}
\bar{R}_{k,i}^{\mathrm U}[n] &=  W {\alpha_{k,i}^{\mathrm I}}[n] \log_2 \bigg( 1+ \frac{\gamma_{k,i}^{\mathrm{IJU}}[n]}{u_k[n]} \bigg), \label{eqn:data_rate_in_sub2} \\
\gamma_{k,i}^{\mathrm{IJU}}[n] &= \frac{p_{k,i}^{\mathrm I}[n] \beta_0} {\text{Tr}(\mathbf{H}_k^{\mathrm{JU}}[n] \mathbf{Z}_i^{\mathrm J}[n]) + W N_0}, \\
\bar{P}_{\mathrm{total}}^{\mathrm I}[n] &= \sum_{k=1}^{K} \sum_{i=1}^{N_{\mathrm{F}}} \alpha_{k}^i[n] p_{k}^i[n] + P_{\mathrm{C}}+ \bar{P}_{\mathrm{flight}}^{\mathrm I}[n], \label{eqn:equivalent_P_bar_total_I} \\
\bar{P}_{\mathrm{flight}}^{\mathrm I}[n] &=  P_o \bigg( 1 + \frac{3 \|\mathbf{v}^{\mathrm I}[n]\| ^2}{\Omega^2 r^2} \bigg) + \frac{P_i v_0}{\upsilon^{\mathrm I}[n]} \notag \\
&+  \frac{1}{2}d_0\rho sA\|\mathbf{v}^{\mathrm I}[n]\|^3, \,\, \text{and} \\
\gamma_{k,e,i}^{\mathrm{IJE}}[n] &= \frac{p_{k,i}^{\mathrm I}[n] \beta_0} {\text{Tr}(\mathbf{H}_e^{\mathrm{JE}}[n] \mathbf{Z}_i^{\mathrm J}[n]) + W N_0}.
\end{align}

Note that $\bar{R}_{k,i}^{\mathrm U}[n]$ and $\bar{P}_{\mathrm{flight}}^{\mathrm I}[n]$ are convex w.r.t. $u_k[n]>0$ and $\upsilon^{\mathrm I}[n]>0$, respectively.
It can be proved that the problems in \eqref{opt_prob:sub2_origin} and \eqref{opt_prob:sub2_bar}
are equivalent as inequality constraints $\mathrm{C22}$ and $\mathrm{C23}$ are always satisfied with equality at the optimal solution of \eqref{opt_prob:sub2_bar}.
Then, we handle the location uncertainty of eavesdropper $e$ by rewriting constraint $\overline{\mathrm{C7}}$ as: 
\begin{eqnarray} \label{eqn:location_uncertainty_constrain_C7_bar_rewrite}
\underset{ \| \Delta \mathbf{t}_e^{\mathrm{E}}\| \leq Q_e^{\mathrm{E}} }{\max}  - \| \hat{\mathbf{t}}_e^{\mathrm{E}} + \Delta \mathbf{t}_e^{\mathrm{E}} - \mathbf{t}^{\mathrm I}[n]\|^2 - H^2 + \frac{\gamma_{k,e,i}^{\mathrm{IJE}}[n]} {\Gamma_{\mathrm{th}}} \leq 0.
\end{eqnarray}
Note that the location uncertainty introduces an infinite number of constraints in $\overline{\mathrm{C7}}$.
To circumvent this difficulty, we apply the $\mathcal{S}$-Procedure \cite{cui2018robust} and transform $\overline{\mathrm{C7}}$ into a finite number of linear matrix inequalities (LMIs) constraints.
In particular,  if there exists a variable $\psi[n] \geq 0$ such that 
\begin{eqnarray} \label{eqn:S_procedure_trans_C7_into_Phi_for_sub2_I}
\Phi (\mathbf{t}^{\mathrm I}[n], \psi[n]) \succeq \mathbf{0}, \forall n,
\end{eqnarray}
holds, where 
\begin{eqnarray}
\Phi (\mathbf{t}^{\mathrm I}[n], \psi[n]) = \left[ \begin{array}{ccc}
(\psi[n] + 1) \mathbf{I}_2 & \mathbf{t}^{\mathrm I}[n] - \hat{\mathbf{t}}_e^{\mathrm E} \\
(\mathbf{t}^{\mathrm I}[n] - \hat{\mathbf{t}}_e^{\mathrm E})^{\mathrm T} & -\psi[n] (Q_e^{\mathrm E})^2 + c[n]
\end{array}
\right]
\end{eqnarray}
and 
\begin{eqnarray}
c[n] &=& \|\mathbf{t}^{\mathrm I}[n]\|^2 - 2 \|(\hat{\mathbf{t}}_e^{\mathrm{E}})^{\mathrm{T}} \mathbf{t}^{\mathrm I}[n]\| + \| \hat{\mathbf{t}}_e^{\mathrm{E}}\|^2 +H^2 \notag \\
&-& \frac{\gamma_{k,e,i}^{\mathrm{IJE}}[n]} {\Gamma_{\mathrm{th}}},
\end{eqnarray}
then the implication \eqref{eqn:S_procedure_trans_C7_into_Phi_for_sub2_I}$\Rightarrow$\eqref{eqn:location_uncertainty_constrain_C7_bar_rewrite} holds.

Next, the non-convexity arises from the numerator of the objective function, constraints $\mathrm{C6}$, $\mathrm{C13}$, and $\mathrm{C23}$ since $\bar{R}_{k,i,\mathrm{lb}}^{\mathrm U} [n]$, $\|\mathbf{t}^{\mathrm I}[n]-\mathbf{t}^{\mathrm J}[n]\|^2$, and $\|\mathbf{v}^{\mathrm I}[n]\|^2$ are convex functions and differentiable w.r.t. $u_k[n]$, $\mathbf{t}^{\mathrm I}[n]$, and $\mathbf{v}^{\mathrm I}[n]$, respectively.
Besides, $c[n]$ in constraint \eqref{eqn:S_procedure_trans_C7_into_Phi_for_sub2_I} is a non-convex function w.r.t. $\mathbf{t}^{\mathrm I}[n]$.
In the following, we aim to establish a lower bound of the objective function and focus on a subset spanned by constraints $\mathrm{C6}$, $\mathrm{C13}$, and $\mathrm{C23}$.
By using the first-order Taylor expansion \cite{EE_fixed_wing} and the SCA \cite{Zhang2018Securing,wei2018multi},
for a given feasible solution $u_k^{(j^{\mathrm{A3}})}[n]$, $(\mathbf{t}^{\mathrm I}[n])^{j^{\mathrm{A3}}}$, and $(\mathbf{v}^{\mathrm I}[n])^{j^{\mathrm{A3}}}$, we have inequalities \eqref{eqn:Rate_lb_in_Sub2_Algo2}, \eqref{eqn:t_I_sca}, and \eqref{eqn:v_I_sca} at the top of next page, respectively.
\begin{figure*}
\begin{align} \label{eqn:Rate_lb_in_Sub2_Algo2}
\bar{R}_{k,i}^{\mathrm U}[n] &\geq (\bar{R}_{k,i,\mathrm{lb}}^{\mathrm U} [n])^{j^{\mathrm{A3}}} =  W {\alpha_{k,i}^{\mathrm I}}[n] \log_2 \bigg( 1+ \frac{\gamma_{k,i}^{\mathrm{IJU}}[n]}{u_k^{(j^{\mathrm{A3}})}[n]} \bigg) - \frac{W {\alpha_{k,i}^{\mathrm I}}[n] \gamma_{k,i}^{\mathrm{IJU}}[n] (u_k[n] - u_k^{(j^{\mathrm{A3}})}[n])}{u_k^{(j^{\mathrm{A3}})}[n] (u_k^{(j^{\mathrm{A3}})}[n] + \gamma_{k,i}^{\mathrm{IJU}}[n]) \ln 2}, \forall k,i,n, \\
\|\mathbf{t}^{\mathrm I}[n]-\mathbf{t}^{\mathrm J}[n]\|^2 &\geq \|(\mathbf{t}^{\mathrm I}[n])^{j^{\mathrm{A3}}} - \mathbf{t}^{\mathrm J}[n]\| ^2 + 2 [(\mathbf{t}^{\mathrm I}[n])^{j^{\mathrm{A3}}}]^{\mathrm{T}}  (\mathbf{t}^{\mathrm I}[n] - (\mathbf{t}^{\mathrm I}[n])^{j^{\mathrm{A3}}}), \, \text{and} \label{eqn:t_I_sca}\\
\|\mathbf{v}^{\mathrm I}[n]\| ^2 &\geq \|(\mathbf{v}^{\mathrm I}[n])^{j^{\mathrm{A3}}}\| ^2 + 2 [(\mathbf{v}^{\mathrm I}[n])^{j^{\mathrm{A3}}}]^{\mathrm{T}}  (\mathbf{v}^{\mathrm I}[n] - (\mathbf{v}^{\mathrm I}[n])^{j^{\mathrm{A3}}}),  \label{eqn:v_I_sca}
\end{align}
\hrulefill
\end{figure*}
Similarly, for a given feasible solution $(\mathbf{t}^{\mathrm I}[n])^{j^{\mathrm{A3}}}$, the following constraint 
\begin{eqnarray} \label{eqn:S_procedure_result_Phi_tilde_for_sub2_I}
\overline{\overline{\mathrm{C7}}}: \widetilde{\Phi}^{(j^{\mathrm{A3}})} (\mathbf{t}^{\mathrm I}[n], \psi[n]) \succeq \mathbf{0}, \forall n,
\end{eqnarray}
where $\widetilde{\Phi}^{(j^{\mathrm{A3}})} (\mathbf{t}^{\mathrm I}[n], \psi[n])$ is given by equation \eqref{equation_phi_tilde} at the top of next page
\begin{figure*}
\begin{eqnarray}
\widetilde{\Phi}^{(j^{\mathrm{A3}})} (\mathbf{t}^{\mathrm I}[n], \psi[n])  = \left[  \begin{array}{ccc}
(\psi[n] + 1) \mathbf{I}_2 & \mathbf{t}^{\mathrm I}[n] - \hat{\mathbf{t}}_e^{\mathrm{E}} \\
(\mathbf{t}^{\mathrm I}[n] - \hat{\mathbf{t}}_e^{\mathrm{E}})^{\mathrm{T}} & - \psi[n] (Q_e^{\mathrm{E}})^2 + \tilde{c}^{(j^{\mathrm{A3}})}[n]
\end{array}
\right] \label{equation_phi_tilde}
\end{eqnarray}
\hrulefill\vspace*{-6mm}
\end{figure*}
and  
\begin{align} \label{eqn:S_procedure_result_Phi_tilde_for_sub2_I_C_tilde}
\tilde{c}^{(j^{\mathrm{A3}})} [n] &= \|\hat{\mathbf{t}}_e^{\mathrm{E}} \|^2 +  2 (\mathbf{t}^{\mathrm I}[n])^{\mathrm{T}} (\mathbf{t}^{\mathrm I}[n])^{j^{\mathrm{A3}}} - ((\mathbf{t}^{\mathrm I}[n])^{j^{\mathrm{A3}}})^2 \notag \\ 
&- 2 (\hat{\mathbf{t}}{_e^{\mathrm E}})^{\mathrm T} \mathbf{t}^{\mathrm I}[n] + H^2 - \frac{\gamma_{k,e,i}^{\mathrm{IJE}}[n]} {\Gamma_{\mathrm{th}}} \leq c[n], \\ 
&\Rightarrow \overline{\mathrm{C7}}. 
\end{align}
Besides, a subset of $\mathrm{C13}$ and $\mathrm{C23}$ is given by
\begin{align}
\overline{\mathrm{C13}}: & \|(\mathbf{t}^{\mathrm I}[n])^{j^{\mathrm{A3}}} - \mathbf{t}^{\mathrm J}[n]\| ^2 + 2 [(\mathbf{t}^{\mathrm I}[n])^{j^{\mathrm{A3}}}]^{\mathrm{T}} \notag \\
& \times (\mathbf{t}^{\mathrm I}[n] - (\mathbf{t}^{\mathrm I}[n])^{j^{\mathrm{A3}}}) \geq d_{\min}^2, \forall n, \\
\overline{\mathrm{C23}}: & \|(\mathbf{v}^{\mathrm I}[n])^{j^{\mathrm{A3}}}\| ^2 + 2 [(\mathbf{v}^{\mathrm I}[n])^{j^{\mathrm{A3}}}]^{\mathrm{T}}  \notag \\
& \times (\mathbf{v}^{\mathrm I}[n] - (\mathbf{v}^{\mathrm I}[n])^{j^{\mathrm{A3}}}) \geq {\upsilon^{\mathrm I}}^2[n], \forall n.
\end{align}

\begin{table}[t] 

\linespread{1}
\begin{algorithm} [H]
\caption{Proposed Algorithm for Solving Sub-problem 2} \label{alg_sub_2}
\begin{algorithmic} [1]

\STATE Initialize the convergence tolerance $\epsilon_3 \rightarrow 0$, the maximum number of iterations for main loop $J^{\mathrm{A3}}_{\max}$, the initial iteration index $j^{\mathrm{A3}} = 1$, and the initial system energy efficiency $q_3^{j^{\mathrm{A3}}} = 0$

\REPEAT[Main Loop: SCA]

\STATE Set $j^{\mathrm{A3}} = j^{\mathrm{A3}} +1$

\STATE Using $\textbf{Algorithm \ref{dinkelbach}}$ with replacing the maximum number of iterations as $J_{\mathrm{in},\max}^{\mathrm{A3}}$, the iteration index as $j_{\mathrm{in}}^{\mathrm{A3}}$, the initial system energy efficiency as $q_3^{(j_{\mathrm{in}}^{\mathrm{A3}})}$, variables as $\{\underline{\mathcal{T_I}}^{(j_{\mathrm{in}}^{\mathrm{A3}})}$, $\underline{\mathcal{V_I}}^{(j_{\mathrm{in}}^{\mathrm{A3}})}$, $\underline{\mathcal{U_K}}^{(j_{\mathrm{in}}^{\mathrm{A3}})}$, $\underline{\Upsilon_{\mathcal I}}^{(j_{\mathrm{in}}^{\mathrm{A3}})}\}$, the total achievable data rate function as $\bar{R}(\underline{\mathcal{U_K}}^{(j_{\mathrm{in}}^{\mathrm{A3}})})$, and the total power consumption as $\bar{P}(\underline{\mathcal{V_I}}^{(j_{\mathrm{in}}^{\mathrm{A3}})}, \underline{\Upsilon_{\mathcal I}}^{(j_{\mathrm{in}}^{\mathrm{A3}})})$ to
obtain $\{\underline{\mathcal{T_I}}^{(j^{\mathrm{A3}})}$, $\underline{\mathcal{V_I}}^{(j^{\mathrm{A3}})}$, $\underline{\mathcal{U_K}}^{(j^{\mathrm{A3}})}$, $\underline{\Upsilon_{\mathcal I}}^{(j^{\mathrm{A3}})}\}$ and $q_3^{(j^{\mathrm{A3}})}$

\UNTIL{$j^{\mathrm{A3}} = J^{\mathrm{A3}}_{\max}$ or $\frac{|q_3^{(j^{\mathrm{A3}})} - q_3^{(j^{\mathrm{A3}}+1)}|}{q_3^{(j^{\mathrm{A3}})}} \leq \epsilon_3$}
\STATE Return $\{\mathcal{T_I}^*$, $\mathcal{V_I}^*$, $\mathcal{U_K}^*$, $\Upsilon_{\mathcal I}^*\}$ = $\{\underline{\mathcal{T_I}}^{(j^{\mathrm{A3}})}$, $\underline{\mathcal{V_I}}^{(j^{\mathrm{A3}})}$, $\underline{\mathcal{U_K}}^{(j^{\mathrm{A3}})}$, $\underline{\Upsilon_{\mathcal I}}^{(j^{\mathrm{A3}})}\}$ and $q_3^* = q_3^{(j^{\mathrm{A3}})}$

\end{algorithmic}
\end{algorithm}
\vspace*{-15mm}
\end{table}

Now, we obtain a lower bound of the objective function via replacing the denominator and the numerator of the original objective function in \eqref{opt_prob:sub2_bar} by its equivalent form in \eqref{eqn:equivalent_P_bar_total_I} and the lower bound of average total data rate in \eqref{eqn:Rate_lb_in_Sub2_Algo2}, respectively.
Besides, we replace constraints $\mathrm{C13}$ and $\mathrm{C23}$ by $\overline{\mathrm{C13}}$ and $\overline{\mathrm{C23}}$, respectively.
Therefore, we can obtain a suboptimal solution of \eqref{opt_prob:sub2_bar} via solving the following optimization problem: 
\begin{align} \label{opt_prob:sub2_lb_iterate}
\underset {\mathcal{T_I},\mathcal{V_I},\,\mathcal{U_K},{\bm\Upsilon_{\mathcal I}},{\bm \Psi}} {\text{maximize}} &\,\, \frac{\sum_{n=1}^{N} \sum_{k=1}^{K} \sum_{i=1}^{N_{\mathrm{F}}} (\bar{R}_{k,i,\mathrm{lb}}^{\mathrm U} [n])^{j^{\mathrm{A3}}}} {\sum_{n=1}^{N} (\bar{P}_{\mathrm{total}}^{\mathrm I}[n] + P_{\mathrm{total}}^{\mathrm J}[n])} \\
\mathrm{s.t.}\,\, &\,\, \overline{\overline{\mathrm{C7}}},\mathrm{C8} - \mathrm{C12},\overline{\mathrm{C13}}, \mathrm{C22},\overline{\mathrm{C23}}, \mathrm{C24}, \notag \\
\overline{\mathrm{C5a}}: &\,\, \bar{P}_{\mathrm{total}}^{\mathrm I}[n] \leq P_{\max}^{\mathrm I}, \forall n, \notag \\
\overline{\overline{\mathrm{C6}}} : &\,\, \frac{1}{N} \sum_{n=1}^{N} \sum_{i=1}^{N_{\mathrm{F}}} (\bar{R}_{k,i,\mathrm{lb}}^{\mathrm U} [n])^{j^{\mathrm{A3}}} \geq  R_{\min}, \forall k, \notag \\
\mathrm{C25}: &\,\, \psi[n] \geq 0, \forall n, \notag 
\end{align}
where $\bm{\Psi}=\{\psi[n], \forall n\}$.
Note that a solution satisfies the constraints in \eqref{opt_prob:sub2_lb_iterate} would satisfy the one in \eqref{opt_prob:sub2_bar}.
Now, similar to the approach for solving sub-problem 1, we apply the Dinklebach's method for a given $\{(\mathbf{t}^{\mathrm I}[n])^{j^{\mathrm{A3}}}$, $(\mathbf{v}^{\mathrm I}[n])^{j^{\mathrm{A3}}}\}$ and $q_3^{(j^{\mathrm{A3}})}$, we solve the following convex optimization problem iteratively\footnote{The problem in \eqref{opt_prob:sub2_final} can be easily solved by dual decomposition or numerical convex program solvers.}: 
\begin{align} \label{opt_prob:sub2_final} &\{\underline{\mathcal{T_I}},\underline{\mathcal{V_I}},\underline{\mathcal{U_K}},\underline{\Upsilon_{\mathcal I}}\} \notag \\
& =  \arg \, \underset {\mathcal{T_I},\mathcal{V_I},\,\mathcal{U_K},{\bm \Psi},{\bm\Upsilon_{\mathcal I}}} {\text{maximize}} \,\, \sum_{n=1}^{N} \sum_{k=1}^{K} \sum_{i=1}^{N_{\mathrm{F}}} (\bar{R}_{k,i,\mathrm{lb}}^{\mathrm U} [n])^{j^{\mathrm{A3}}} \notag \\
&- q_3^{(j_{\mathrm{in}}^{\mathrm{A3}})} \sum_{n=1}^{N} (\bar{P}_{\mathrm{total}}^{\mathrm I}[n] + P_{\mathrm{total}}^{\mathrm J}[n]) \\
\mathrm{s.t.}\,\, &\,\, \overline{\mathrm{C5a}},\overline{\overline{\mathrm{C6}}},\overline{\overline{\mathrm{C7}}},\mathrm{C8} - \mathrm{C12},\overline{\mathrm{C13}}, \mathrm{C22}, \overline{\mathrm{C23}}, \mathrm{C24}, \mathrm{C25}, \notag
\end{align}
where $\{\underline{\mathcal{T_I}},\underline{\mathcal{V_I}},\underline{\mathcal{U_K}},\underline{\Upsilon_{\mathcal I}}\}$ is the optimal solution of \eqref{opt_prob:sub2_final} for a given $q_3^{(j_{\mathrm{in}}^{\mathrm{A3}})}$.
The problem optimization in \eqref{opt_prob:sub2_final} is a convex formulation which can be easily solved by CVX \cite{cvx}.
The proposed algorithm for solving sub-problem 2 is summarized in $\textbf{Algorithm \ref{alg_sub_2}}$.

\subsection{Overall Algorithm}

\begin{table}[t] 

\linespread{1}
\begin{algorithm} [H]
\caption{Overall Algorithm for Solving Problem \eqref{opt_prob:overall}} \label{alg_total}
\begin{algorithmic} [1]

\STATE Initialize the convergence tolerance $\epsilon_4 \rightarrow 0$, the maximum number of iterations $J_{\max}^{\mathrm{A4}}$, the initial iteration index $j^{\mathrm{A4}} = 1$, and the initial trajectory $\{\mathbf{t}^{\mathrm I}[n], \mathbf{v}^{\mathrm I}[n],\mathbf{t}^{\mathrm J}[n], \mathbf{v}^{\mathrm J}[n]\}$

\REPEAT

\STATE Set $j^{\mathrm{A4}} = j^{\mathrm{A4}}+1$

\STATE Using $\textbf{Algorithm \ref{alg_sub_1}}$ obtain the suboptimal result $q_1$, $\{\alpha_k^i[n], p_k^i[n],\mathbf{Z}_i^{\mathrm J}[n]\}$

\STATE Using $\textbf{Algorithm \ref{alg_sub_2}}$ obtain the suboptimal result $q_3$,  $\{\mathbf{t}^{\mathrm I}[n], \mathbf{v}^{\mathrm I}[n]\}$

\UNTIL{$j^{\mathrm{A4}}=J_{\max}^{\mathrm{A4}}$ or $\frac{|q_3^{(j^{\mathrm{A4}})} - q_3^{(j^{\mathrm{A4}}+1)}|}{q_3^{(j^{\mathrm{A4}})}} \leq \epsilon$}

\RETURN $ {\alpha_k^i}^*[n]=\alpha_k^i[n], {p_k^i}^*[n]=p_k^i[n], {\mathbf{Z}_i^{\mathrm J}}^*[n]=\mathbf{Z}_i^{\mathrm J}[n], {\mathbf{t}^{\mathrm I}}^*[n]=\mathbf{t}^{\mathrm I}[n], {\mathbf{v}^{\mathrm I}}^*[n]=\mathbf{v}^{\mathrm I}[n]$, and $q^*=q_3^{(j^{\mathrm{A4}})}$

\end{algorithmic}
\end{algorithm}
\end{table}

The overall proposed iterative algorithms for solving the two sub-problems \eqref{opt_prob:sub1_origin} and \eqref{opt_prob:sub2_origin} are summarized in $\textbf{Algorithm \ref{alg_total}}$.
Since the feasible solution set of \eqref{opt_prob:overall} is compact and its objective value is non-decreasing over iterations via solving the sub-problem in \eqref{opt_prob:sub1_origin} and \eqref{opt_prob:sub2_origin}, iteratively, the solution of the proposed algorithm is guaranteed to converge \cite{Alternating}.
Since we handle the problem with SCA and $\mathcal{S}$-Procedure, the obtained solution converges to a suboptimal optimal solution {\cite{Alternating,bezdek2002some,wu2018joint,opial1967weak,5519540}} of the original problem in \eqref{opt_prob:overall}.

On the other hand, as the computational complexity of solving sub-problem 1 is dominated by the semidefinite programming (SDP), the computational complexity of the proposed suboptimal algorithm is given by equation \eqref{equation_computational_complexity} at the top of this page\cite{nesterov1994interior,nonlinear}. 
\begin{figure*}
\begin{eqnarray}
&\mathcal{O} \bigg( J_{\max}^{\mathrm{A4}} \bigg( \underset{\text{Sub-problem 1}}{\underbrace{(\mathcal{M}_1 \mathcal{N}_1^3 + \mathcal{M}_1^2 \mathcal{N}_1^2 + \mathcal{M}_1^3 \mathcal{N}_1) \times J_{\max}^{\mathrm{A1}} J_{\mathrm{in},\max}^{\mathrm{A2}} \bigg(\sqrt{\mathcal{N}_1} \log \bigg(\frac{1}{\Delta_1}\bigg) \bigg)}} \notag \\
&+ \underset{\text{Sub-problem 2}} {\underbrace{\mathcal{M}_2 \mathcal{N}_2^2 \times J_{\max}^{\mathrm{A3}} J_{\mathrm{in},\max}^{\mathrm{A3}} \bigg(\sqrt{\mathcal{N}_2} \log \bigg(\frac{1}{\Delta_2} \bigg) \bigg) }} \bigg) \bigg), \label{equation_computational_complexity}
\end{eqnarray}
\hrulefill
\end{figure*}
Note that $\mathcal{M}_1 = 10 NKN_{\mathrm F} + NKEN_{\mathrm F} + 2NN_{\mathrm F} + 4N + K$, $\mathcal{N}_1 = 3 NKN_{\mathrm F} + N_{\mathrm J}^2NN_{\mathrm F} + N_{\mathrm J}^2NKN_{\mathrm F}$, $\mathcal{M}_2 = 9N + NK + K$, and $\mathcal{N}_2 = 4N + NK$.
Besides, $\Delta_1>0$, and $\Delta_2>0$ denote the solutions of the sub-problem 1 and sub-problem 2, respectively.
We note that the proposed suboptimal algorithm has a polynomial time computational complexity.

\section{Numerical Results}

\begin{table}[t] 
\scriptsize
\caption{Simulation parameters \cite{EE_fixed_wing,8408554,rotary_wing_power,8589002}.} \label{simulation_setting}
\begin{center}
\begin{tabular}{ c | c | c | c }
  \hline			
  Notations                     & Simulation value              & Notations                 & Simulation value \\ \hline
  $\Omega$                      & 300 radians/second            & $K$                       & 2  \\
  $r$                           & 0.4 meter                     & $E$                       & 2  \\
  $\rho$                        & 1.225 $\mathrm{kg/m^3}$       & $\tau$                    & 0.1 s  \\
  $s$                           & 0.05                          & $N_{\mathrm{F}}$          & 128  \\
  $A_{\mathrm r}$               & 0.503 $\mathrm{m^2}$          & $B$                       & 1 MHz  \\
  $P_o$                         & 79.86 W                       & $W$                       & 7.8 kHz \\
  $P_i$                         & 88.63 W                       & $N_0$                     & -160 dBm/Hz \\
  $v_0$                         & 4.03 m/s                      & $P^{\mathrm I}_{\mathrm{C}}$  & 30 dBm  \\
  $d_0$                         & 0.3                           & $P^{\mathrm J}_{\mathrm{C}}$  & 30 dBm  \\
  $V^{\mathrm I}_{\max}$        & 30 m/s                        & $\zeta^{\mathrm I}$       & 2   \\
  $V^{\mathrm I}_{\mathrm{acc}}$ & 4 m/$\text{s}^{\text 2}$     & $\zeta^{\mathrm J}$       & 2  \\
  $P^{\mathrm I}_{\max}$        & 65 dBm                        & $\lambda_{\mathrm c}$     & $10^{-10}$ m  \\
  $P^{\mathrm J}_{\max}$        & 65 dBm                        & $\Delta_{\mathrm J}$      & 0.1 m    \\
  $N_{\mathrm Jx}$              & 5                             & $R_{\min}$                & 6 Mbits/s  \\
  $N_{\mathrm Jy}$              & 5                             & $\Gamma_{\mathrm{th}}$    & $10^{-3}$ bps/subcarrier  \\
  $\mathbf{t}_0$                & $[0,0]$ m                     & $\mathbf{t}_1^{\mathrm U}$ & $[350,100]$ m  \\
  $\mathbf{t}_{\mathrm{F}}$     & $[500,500]$ m                 & $\mathbf{t}_2^{\mathrm U}$ & $[150,400]$ m  \\
  $\hat{\mathbf{t}}_1^{\mathrm E}$  & $[400,100]$ m         & $Q_e^{\mathrm E}$         & $[71,141]$ m \\
  $\hat{\mathbf{t}}_2^{\mathrm E}$  & $[250,250]$ m         & $H$                       & 100 m \\
  $P^{\mathrm{I}}_{\mathrm{peak}}$  & 30 dBm                & $d_{\min}$                & 1 m \\
  $P^{\mathrm{J}}_{\mathrm{peak}}$  & 30 dBm                & $J_{\max}^{\mathrm{A1}}$  & 10 \\
  $J_{\max}^{\mathrm{A3}}$          & 10                    & $J_{\max}^{\mathrm{A4}}$  & 5 \\
  $J_{\mathrm{in},\max}^{\mathrm{A3}}$  & 10 \\
  \hline
\end{tabular}
\end{center}
\end{table}

In this section, we evaluate the performance of the proposed algorithm via simulations.
The simulation setups are summarized in Table \ref{simulation_setting}.
In our simulations, we compare the system energy efficiency of the proposed algorithm ``PA'', with the other three baseline schemes:
(a) \emph{No jammer UAV (NJ)}, which has only an information UAV in this scheme.
The suboptimal resource allocation and UAV's trajectory for ``NJ'' can be obtained by using a similar approach as in our previous work \cite{cai2019energy}.
(b) \emph{Single-antenna jammer UAV (SAJ)}, in which both the information UAV and jammer UAV are all equipped with a single-antenna to provide secure communication.
Since the problem formulation of the ``PA'' subsumes ``SAJ'', the system performance of ``SAJ'' can be achieved by solving the designed problem with ``PA'' and setting the number of antenna array $N_{\mathrm Jx} = N_{\mathrm Jy} =1$;
{(c) \emph{Zero-acceleration information UAV (ZAI)}, where the information UAV's flight velocity remains unchanged but is optimized by our proposed scheme;}
(d) \emph{Straight locus information UAV (SLI)}, where the information UAV cruises with a straight locus trajectory from the initial point to the final point with a constant speed and the jammer UAV has the same setting as in the ``PA''.
Since ``SLI'' is another subcase of the problem formulation for ``PA'', the suboptimal solution can be obtained by optimizing resource allocation with fixing the information UAV's trajectory.
{Since the initial information-UAV trajectory will affect the performance of the proposed suboptimal solution, we have tried different reasonable trajectories as an initial point, e.g.,
(1) \emph{Straight forward flight from the initial point to the destination} (SFF);
(2) \emph{A path passing through all users' location once};
(3) \emph{A path along the boundary of the service area}, and found out that ``SFF'' provides the best performance.
As a result, in the simulation section, we adopt ``SFF'' as the initial trajectory for the proposed algorithm.}

\subsection{Convergence of the Proposed Algorithm and Baseline Schemes}

\begin{figure*} 
\centering
\begin{minipage}{0.49\textwidth}
    \centering
    \includegraphics[width=3.5 in] {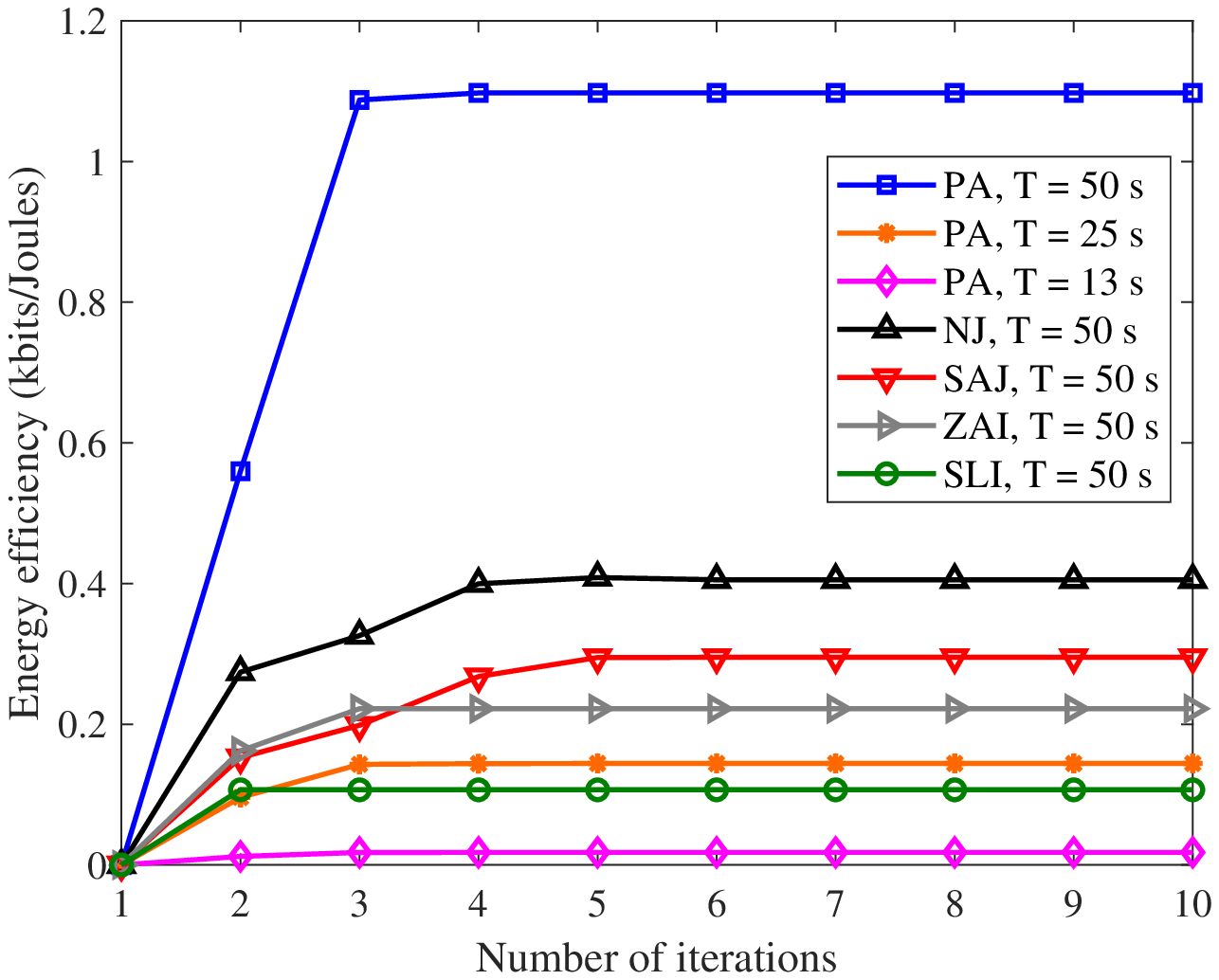} 
    \caption{Energy efficiency versus the number of iterations.}
    \label{EE_Vs_iteration}
\end{minipage}
\begin{minipage}{0.49\textwidth}
    \centering
    \includegraphics[width=3.5 in]{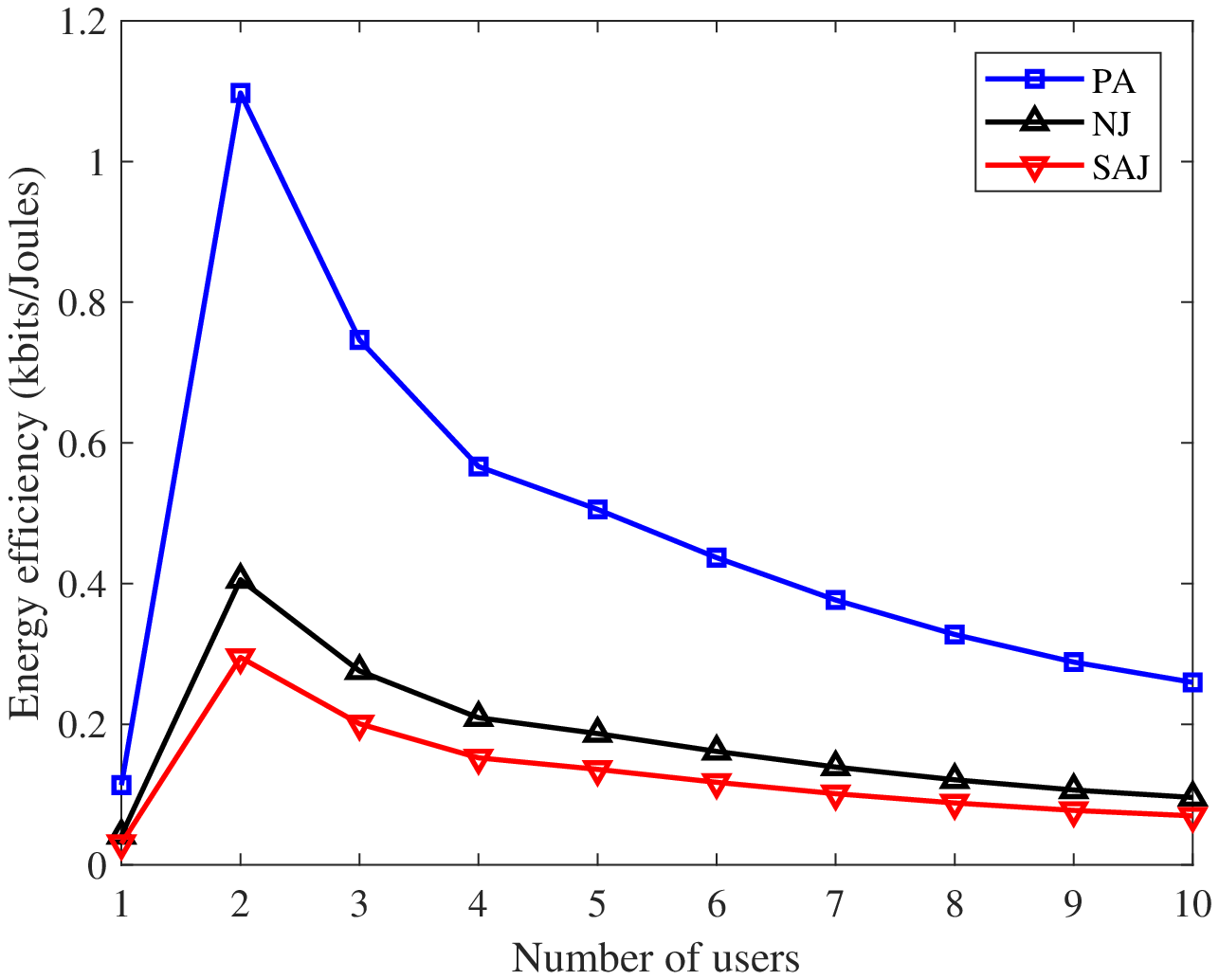} 
    \caption{Energy efficiency versus the number of users.}
    \label{EE_vs_number_of_user}
\end{minipage}
\end{figure*}

Figure \ref{EE_Vs_iteration} illustrates the convergence behavior of the alternating optimization $\textbf{Algorithm \ref{alg_total}}$ for the maximization of the system energy efficiency.
We compare the system energy efficiency of our proposed scheme for three different mission time durations, $T=50$ s, $T=25$ s, and $T=13$ s, which correspond to the number of time slot $N=500$, $N=200$, and $N=130$, respectively.
The jammer UAV orbits around the \emph{center of the eavesdroppers areas (CEA)}\cite{wu2018joint}, as shown in Figure \ref{different_jammer}.
It can be seen from Figure \ref{EE_Vs_iteration} that the system energy efficiency of the proposed scheme with different $T$ converges to the corresponding suboptimal solutions within only $5$ iterations which demonstrates the fast convergence of the proposed alternating optimization algorithm.
Thus, in the following simulations, we set the maximum number of iterations as $5$ to illustrate the performance of the proposed algorithm.
For comparison, we also demonstrate the convergence behavior of four baseline schemes ``NJ'', ``SAJ'', ``ACS'', and ``SLI'' while the mission time duration for baseline schemes is fixed as $T=50$ s, their performance and corresponding trajectory will be discussed in the following.

\subsection{Impact of Number of Users}

{In order to show the impact of the number of users, $K$, on the system performance, we vary the number of users, from 1 to 9, and the location of these users in x-dimension and y-dimension are given by $x_k^{\mathrm U}$ $ = [300;200;100;300;$ $500;900;700;$ $300;500;100]$ and $y_k^{\mathrm U}$ $ = [800;700;100;$ $300;800;900;$ $700;200;500;300]$, respectively.
The minimum data rate requirement for each user is set $R_{\min} = 1$ Mbits/s in this simulation.
Other setups remain the same as before.
The corresponding system energy efficiency versus number of users is shown in Figure \ref{EE_vs_number_of_user}.
We can observe that for all the mentioned schemes, the energy efficiency achieved with $K = 2$ is much higher than that with $K = 1$.
In fact, when the number of users is small, the UAV can exploit the multiuser diversity via the proposed scheduling for improving the system performance.
However, when there are more than 2 users, the minimum data rate constraints $\mathrm{C6}$ become stringent and the resource allocator becomes less flexible in optimizing the usage of system resources leading to the decrease of system energy efficiency.
Besides, the system performance of ``PA'' is always better than that of other baseline schemes while increasing the number of user $K$.}

\subsection{Impact of Jammer UAV's Trajectory}

\begin{figure*}[!t] 
\centering
    \subfigure[The information UAV's trajectories of the proposed algorithm for CSA, CEA, and SFE.]
    {\includegraphics[width=3.5in]{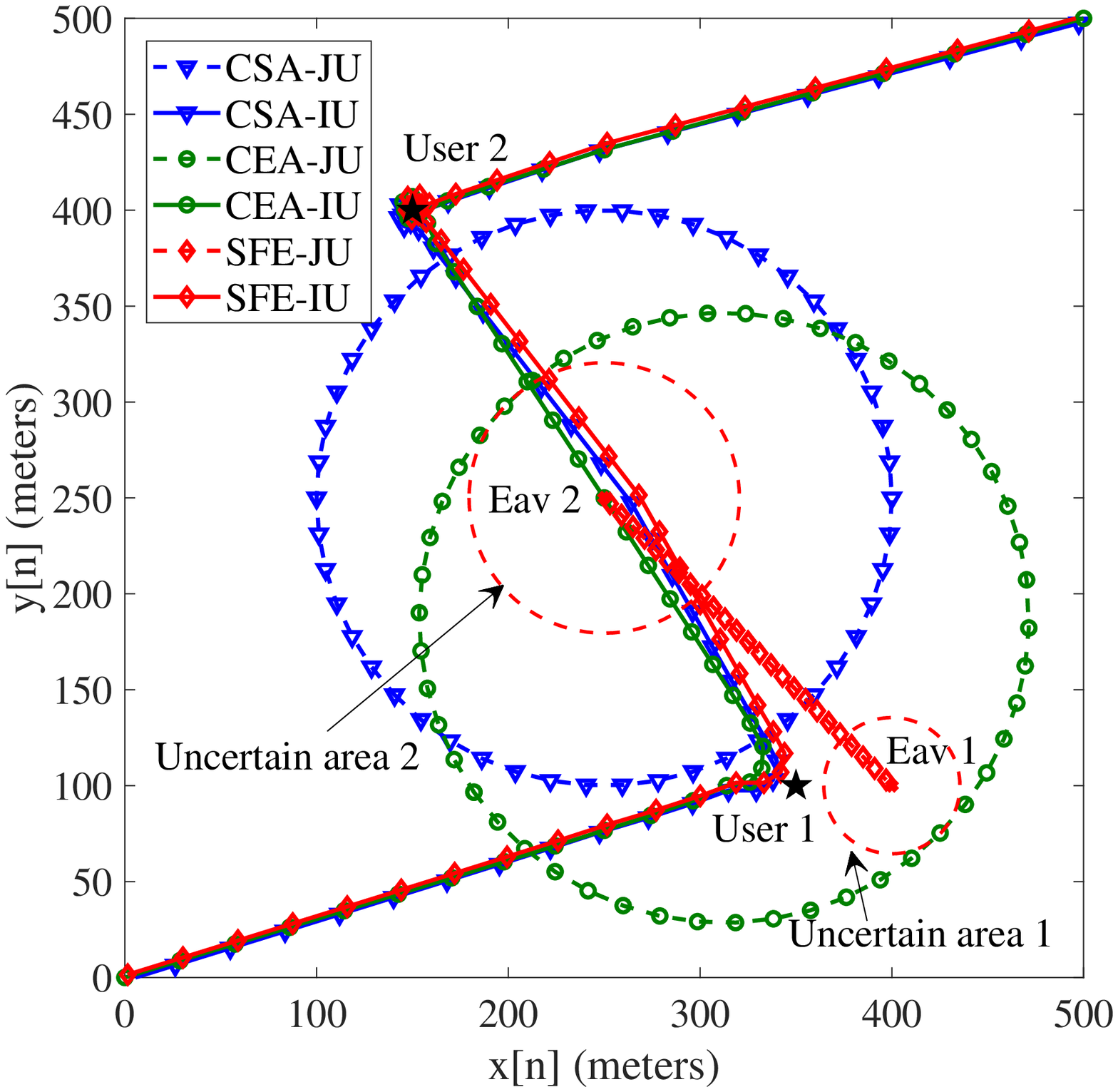}}
    \subfigure[The information UAV's trajectories of the proposed algorithm for CA1, CA2, and CA3.]
    {\includegraphics[width=3.5in]{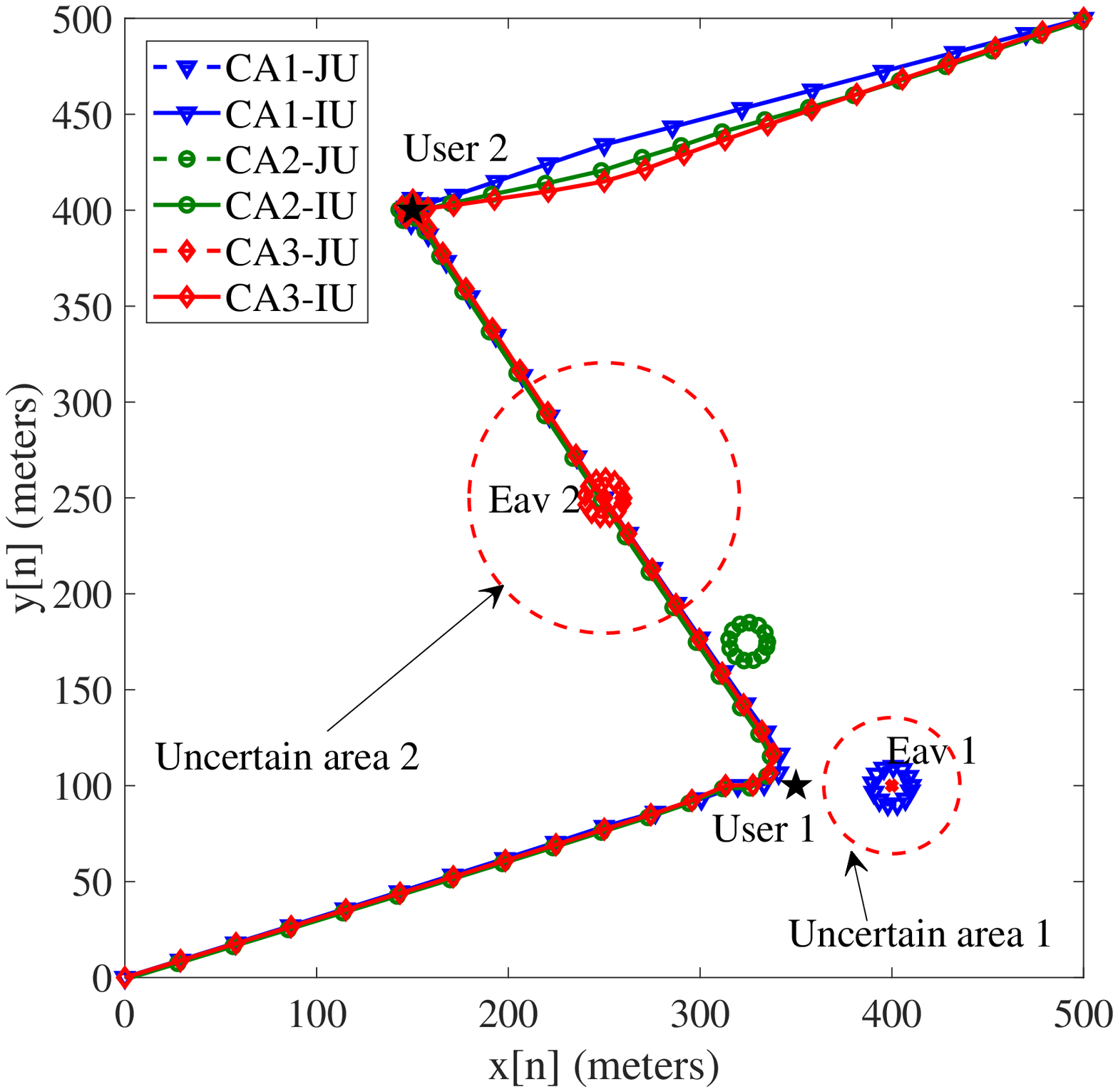}}

\caption{The information UAV's trajectories of the proposed algorithm for different jammer UAV's trajectories.}
\label{different_jammer}
\end{figure*}

Figure \ref{different_jammer} shows the corresponding information UAV's trajectories for different predetermined trajectories of the jammer UAV with $T =50$ s.
{In this paper, we consider six commonly adopted trajectories of the jammer UAV, which have the same flight velocity\footnote{Note that a UAV consumes the minimum flight power when it travels at 10.4 m/s for the considered setting in \cite{rotary_wing_power}.} at 10.4 m/s, with the following proposed scheme.
(a) \emph{Center of the service area (CSA)}, in which the jammer UAV adopts a circular trajectory centered at the center of the service area [250, 250] with a radius of 150 meters \cite{8408554};
(b) \emph{Center of the eavesdroppers area (CEA)}, where the jammer UAV patrols also with a circular trajectory but centered at [312.5, 187.5] (centroid of all estimated eavesdroppers' locations) with a radius of 159 meters \cite{wu2018joint};
(c) \emph{Shuttling flight between the eavesdroppers (SFE)}, where the jammer UAV flight is shuttled back and forth between the estimated locations of the two eavesdroppers during the given time frame;
(d) \emph{Centered at [400, 100] (CA1)}, (e) \emph{Centered at [375, 175] (CA2)}, and (f) \emph{Centered at [250, 250] (CA3)}, in which the jammer UAV has a circular trajectory with a radius of 10 meters centered at the eavesdropper 1's estimated location [400, 100], the middle of two eavesdroppers' estimated locations [375, 175], and the eavesdropper 2's estimated location [250, 250], respectively.
Note that in these schemes, the jammer UAV is equipped with 25 antennas.}
{We can observe in Figure \ref{different_jammer} that by setting a reasonable trajectory of the jammer UAV, e.g., a path cruises among all eavesdroppers, a high system energy efficiency can be achieved compared to the case without jamming UAV.
In fact, the optimized artificial noise would try to compensate the suboptimality caused by the fixed trajectory.
More importantly, the existence of jamming UAV and optimized jamming relieves the security constraint which provides a higher flexibility to the information UAV for adopting an energy efficient short route for communication.
As a result, the information UAV's trajectories are almost the same (with short paths) for different jammer UAV's trajectories.}
This observation will be verified again when we compare our proposed scheme with no jammer in the next section.
Therefore, in the following simulations, we fix the jammer UAV's trajectory as ``CEA'' for illustration.

\subsection{Trajectories of Information UAV}

%
%

\begin{figure*} 
\centering
\begin{minipage}{0.49\textwidth}
    \centering
    \includegraphics[width=3.54 in] {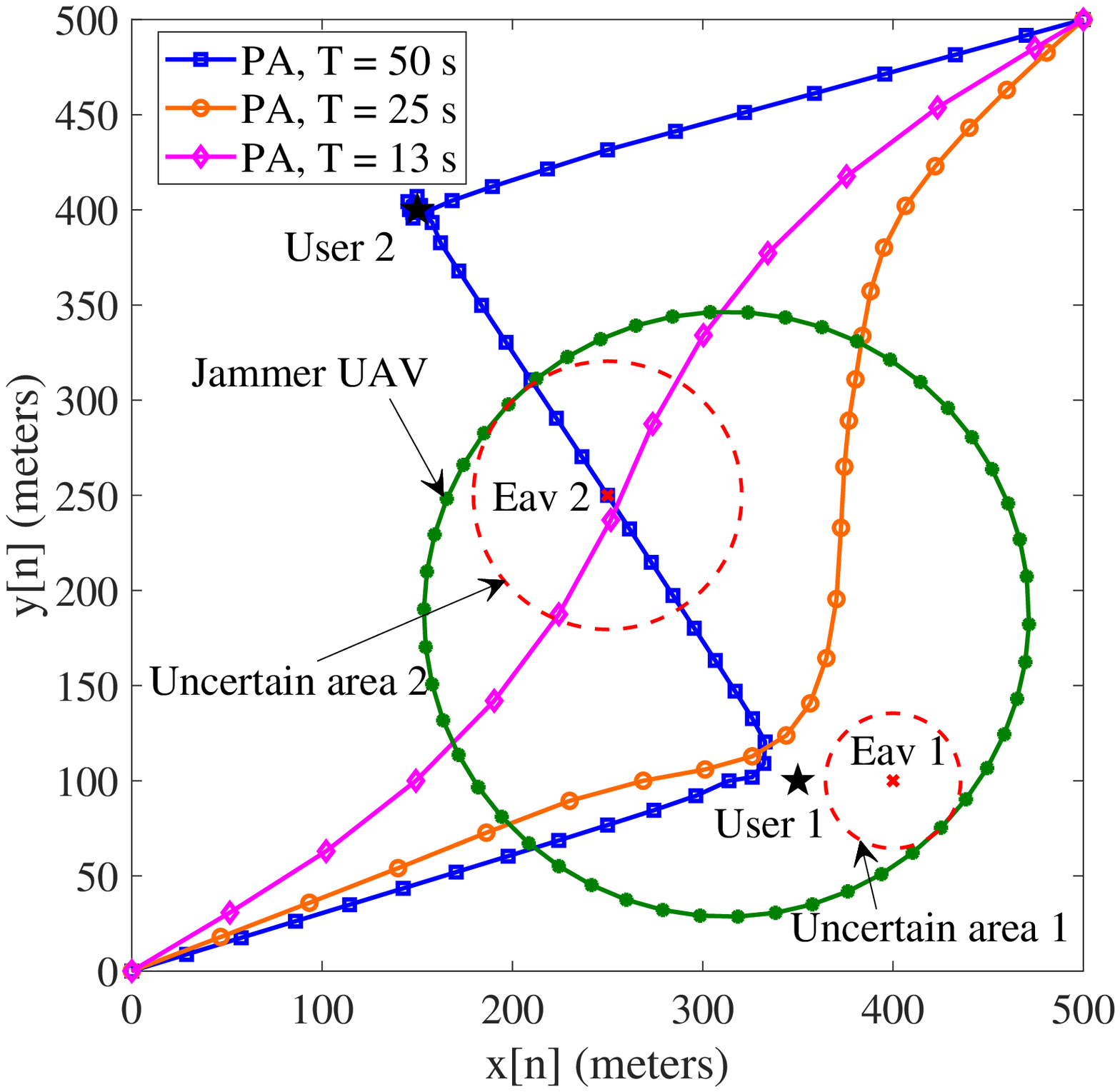} 
    \caption{The UAVs' trajectories of the proposed algorithm with different service time durations.}
  \label{different_duration_time}
\end{minipage}
\begin{minipage}{0.49\textwidth}
    \centering
    \includegraphics[width=3.40 in]{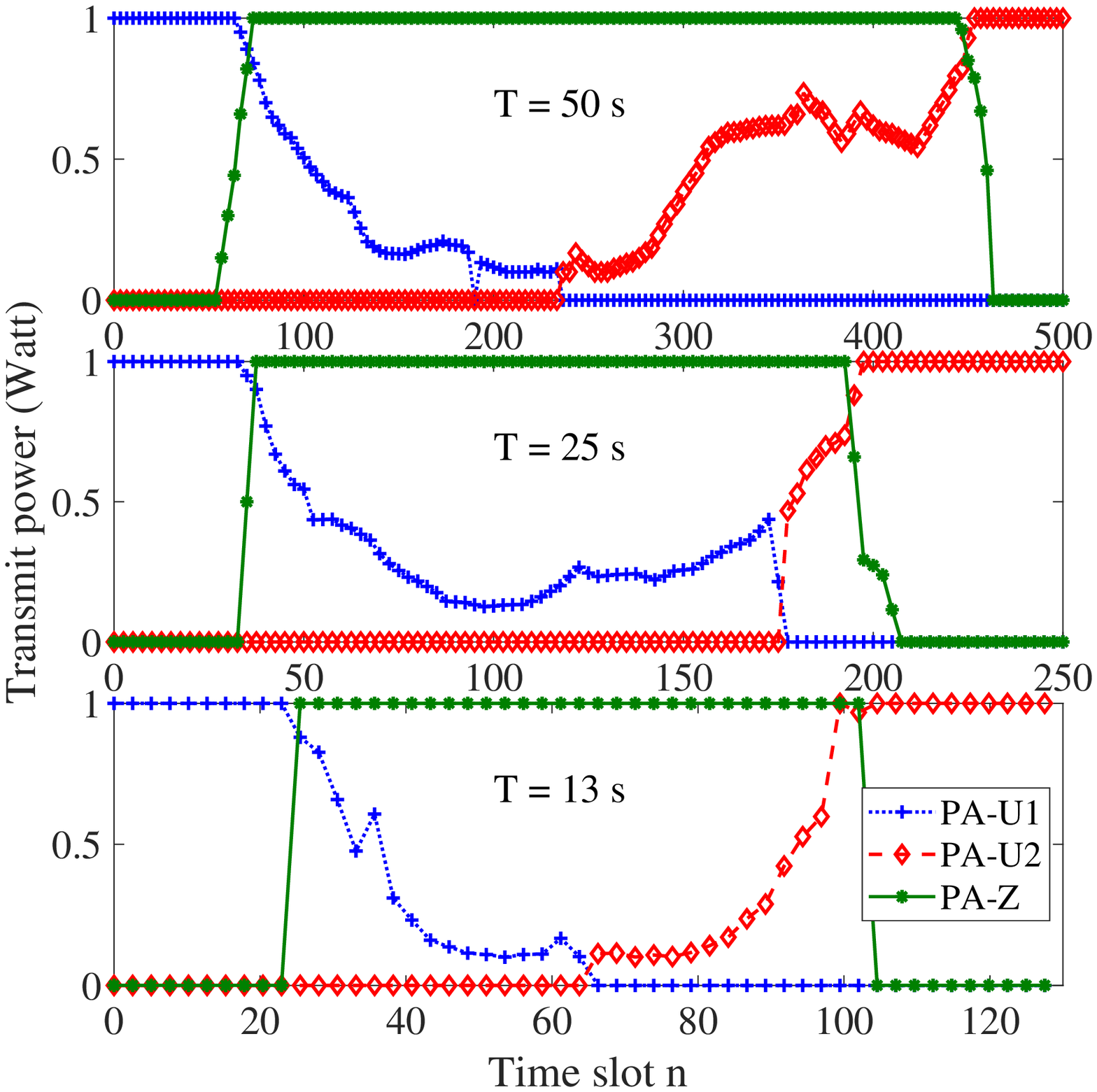} 
  \caption{The communication transmit power to user 1 and user 2 as well as the artificial noise transmit power versus time slots.}
  \label{power_different_duration_time}
\end{minipage}
\end{figure*}

{Figure \ref{different_duration_time} demonstrates the trajectory of the information UAV for the ``PA'' with three different mission time durations, $T=13$ s, $T=25$ s, and $T=50$ s, respectively.
Note that the flight velocity of the information UAV in each time slot can be calculated from the distance between each two adjacent points along its trajectory.
Besides, the corresponding communication transmit power and artificial noise transmit power versus time slots are illustrated in Figure \ref{power_different_duration_time}, where the communication power for user 1, user 2, and the noise power are denoted as ``PA-U1'', ``PA-U2'', and ``PA-Z'', respectively.
Besides, a longer mission completion time enables a higher system energy efficiency for our proposed scheme.
This is because the information UAV's trajectory design becomes more flexible with increasing $T$.
As a result, the mobility of the information UAV can be more efficiently exploited to improve the system energy efficiency.
In the following, for different mission time durations $T$, we will discuss simulation results of the information UAV's trajectory, communication power allocation, and noise power allocation.}

It is observed that when the mission time duration is sufficiently large (e.g., $T=50$ s), the information UAV would maintain a high velocity when it is far away from the users and only fly slowly whenever it is close to any desired user.
This behavior aims to save more time slots for the information UAV to provide high data rate communication when it is close to the users.
Besides, with $T=50$ s, the information UAV would strike a balance between energy consumption and velocity.
In particular, the information UAV hovers above user 2 with the optimized velocity for a long period of time to achieve a high throughput.
In contrast, the information UAV does not hover above user 1 as user 1 is closer to one of the eavesdroppers than user 2 which has a higher potential in information leakage.
Correspondingly, as shown in Figure \ref{power_different_duration_time}, for $T=50$ s, the communication power is allocated solely to user 1 at first half of total time slots, then the remaining time slots are allocated to user 2.
Moreover, when the information UAV is faraway from all the users and eavesdroppers, e.g. at the beginning and ending time slots, the information UAV transmits the highest available communication power and the jammer transmits small power of artificial noise as the leakage SINR of each eavesdropper are relatively small.
However, for those time slots having a high potential of information leakage, not only the jammer UAV transmits the highest artificial noise, but also the information UAV decreases its transmit power to reduce the potential information leakage.
Specifically, by exploiting the spatial degrees of freedom brought by the multiple antennas, the jammer UAV creates a sharp artificial noise beam with full power and steers towards a direction with can impair both eavesdroppers efficiently.
In contrast, when the mission time duration $T$ is 25 s as shown in Figure \ref{different_duration_time}, the information UAV first flies towards to user 1 with a relatively higher velocity then flies slowly to the destination.
{Note that the UAV would slow down but with a reasonable speed when it is close to user 2 instead of stationing since the flight power consumption of the rotary UAV is relatively high when its flight speed is sufficiently low \cite{rotary_wing_power}.}
It can also be observed that the information UAV detours a bit towards user 2 for a more efficient communication.
From Figure \ref{power_different_duration_time}, for $T = 25$ s, the information UAV first communicates with user 1 until the 36-th time slot, where it just crosses outside the locus of the jammer UAV.
Then, the maximum transmit power is allocated solely to user 2 to achieve the minimum data rate requirement.
Additionally, when the total time duration is relatively short (e.g., $T=13$ s), the information UAV flies with the highest speed from the initial point to the final point.
Besides, due to the limited mission completed time, the information UAV flies slightly closer to user 1 at the beginning and later to user 2 for satisfying the individual user's minimum data rate requirement of security communication.
Moreover, the information power allocation and the jamming policy have a similar pattern for the ``PA'' with different total time durations, c.f. Figure \ref{power_different_duration_time}.
These illustrate that the information UAV's trajectory plays an extremely important role in achieving high system energy efficiency and secure communication.

\begin{figure*} 
	\centering
\begin{minipage}{0.49\textwidth}
		\centering
		\includegraphics[width=3.5 in] {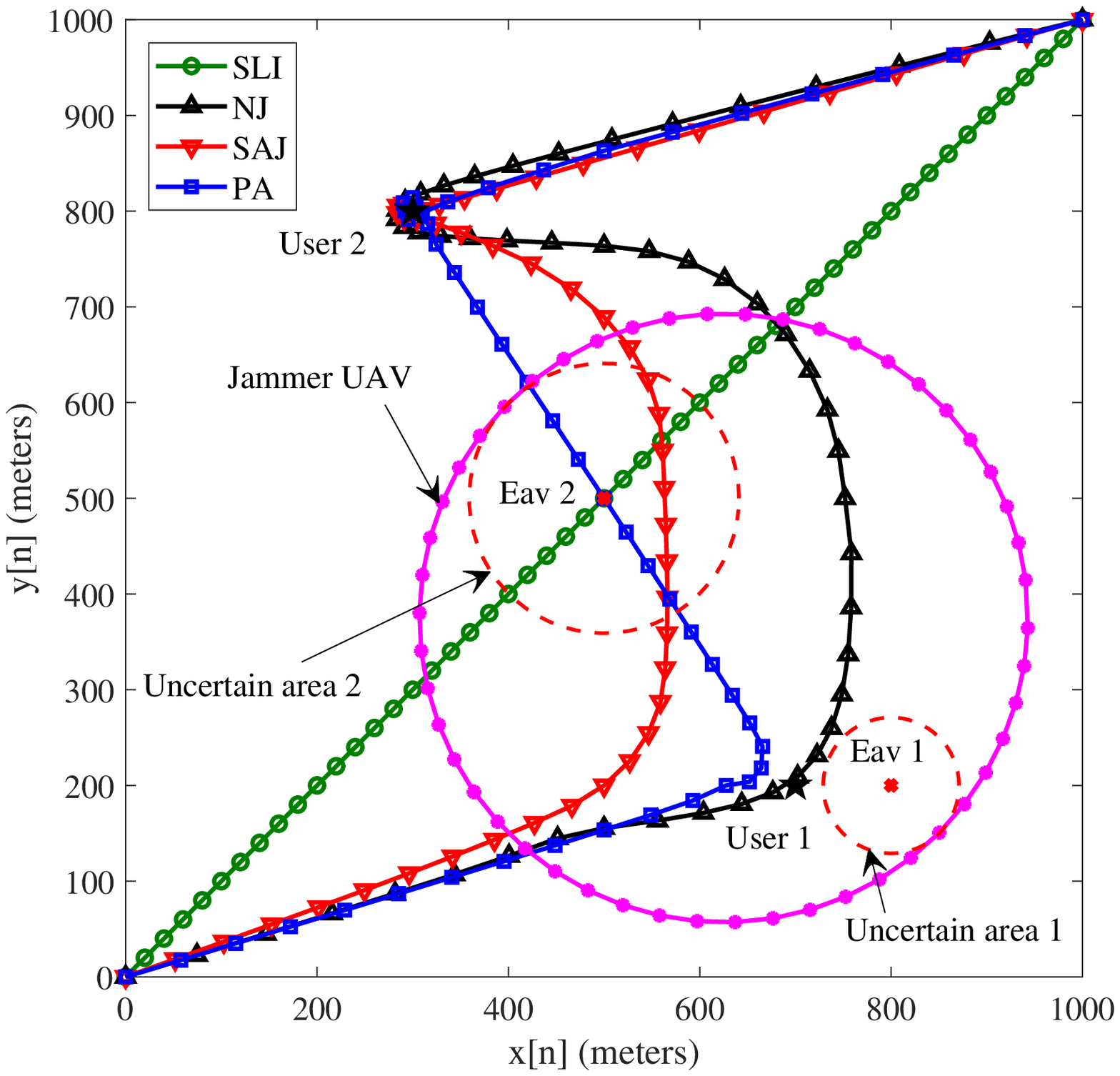} 
  \caption{The UAVs' trajectories of the proposed algorithm and the baseline schemes.}
    \label{trajectory_baseline}
	\end{minipage}
	\begin{minipage}{0.49\textwidth}
		\centering
		\includegraphics[width=3.38 in]{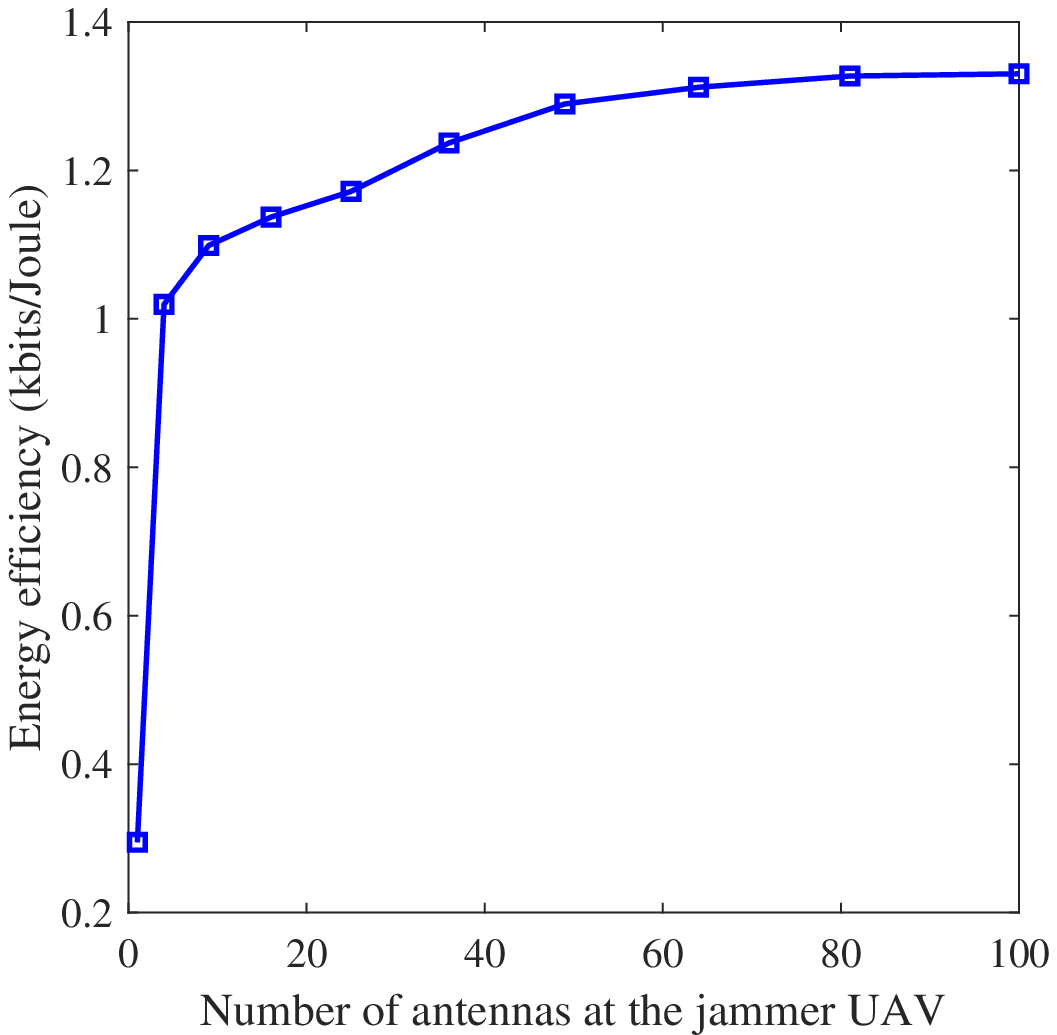} 
  \caption{Energy efficiency versus the number of antennas equipped at the jammer UAV.}
  \label{EE_vs_number_of_antenna}
	\end{minipage}
\end{figure*}
%

\begin{figure}[t] 
  \centering
  \includegraphics[width=3.5 in] {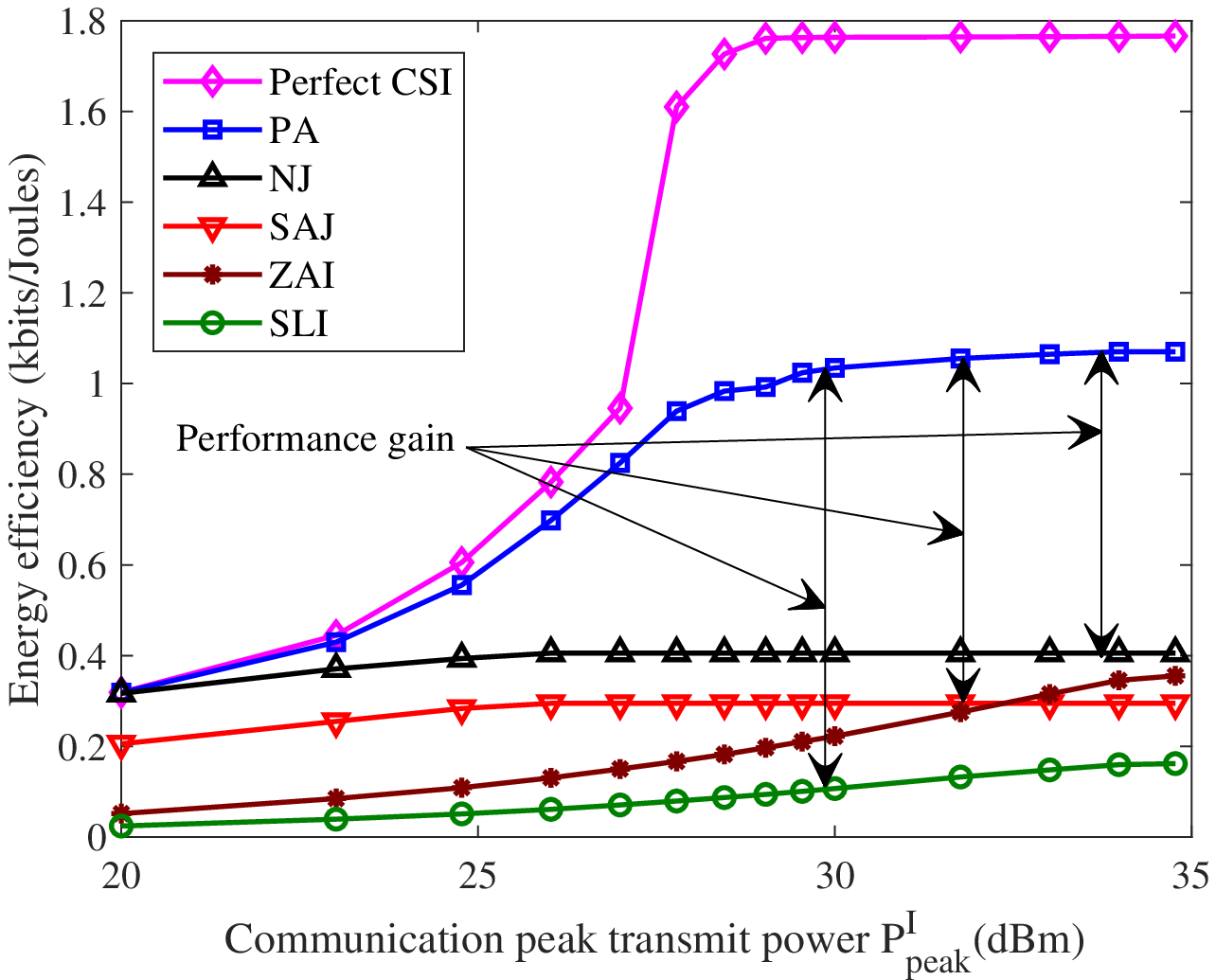} 
  \caption{Energy efficiency versus communication peak transmit power.}
  \label{EE_Vs_P}
\end{figure}

Figure \ref{trajectory_baseline} illustrates the information UAV's trajectories for different schemes, as ``SLI'', ``NJ'', ``SAJ'', and ``PA''.
In this figure, we assume that the mission time duration $T$ is 50 s for all the schemes.
As it can be observed, for ``SLI'', the information UAV flies at a constant speed and following a predefined straight trajectory from the initial point to the destination, which have the lowest energy efficiency in all the considered schemes, c.f. Figure \ref{EE_Vs_iteration} and Figure \ref{EE_Vs_P}.
The information UAV in ``NJ'' scheme first flies towards user 1.
Meanwhile, the information UAV keeps decreasing its transmit power allocated to user 1 for reducing the potential of information leakage.
After passing by user 1, the information UAV starts communicate with user 2 with a small transmit power which adopts an arc trajectory and fly towards user 2.
The detouring trajectory of the information UAV aims to decrease the leakage SINR to eavesdropper 2.
Note that the information UAV only communicates with user 2 with high transmit power when the UAV is far away from eavesdropper 2.
In contrast, the information UAV in ``SAJ'' scheme flies a shorter distance than that of ``NJ'' due to the artificial noise generated by the jammer UAV which relaxes the security requirement on ``SAJ''.
Additionally, comparing all the baseline schemes, the trajectory of information UAV in the ``PA'' does not detour and fly around the uncertain area of the eavesdroppers.
In other words, ``PA'' has a higher flexibility in design the trajectory of the information UAV.
This is a clear evidence of the benefit in utilizing an antenna array at the jammer UAV as it can always focus the artificial noise on the threatened eavesdroppers for guaranteeing secure communication.

\subsection{Energy Efficiency}

{Figure \ref{EE_vs_number_of_antenna} shows the energy efficiency versus the number of antennas equipped at the jammer UAV.
In this simulation, we consider the circuit power consumption for each antenna of the jammer UAV, with $P_{\mathrm{CJ}} = 0.1$ Watt.
It is obviously that the energy efficiency increases with the number of antennas equipped at the jammer UAV as the associated spatial degrees of freedom improve the flexibility in resource allocation.
Besides, the energy efficiency become saturated when the jammer UAV's antenna number is sufficiently large.
This is due to the fact that the circuit power consumption of antennas become a dominate factor in the system performance outweighing the associated performance gain.
In particular, the increase trend of the system energy efficiency presents the contribution of the multiple antennas equipped in the jammer UAV to the system.}

\begin{figure}[t]
  \includegraphics[width=3.5 in]{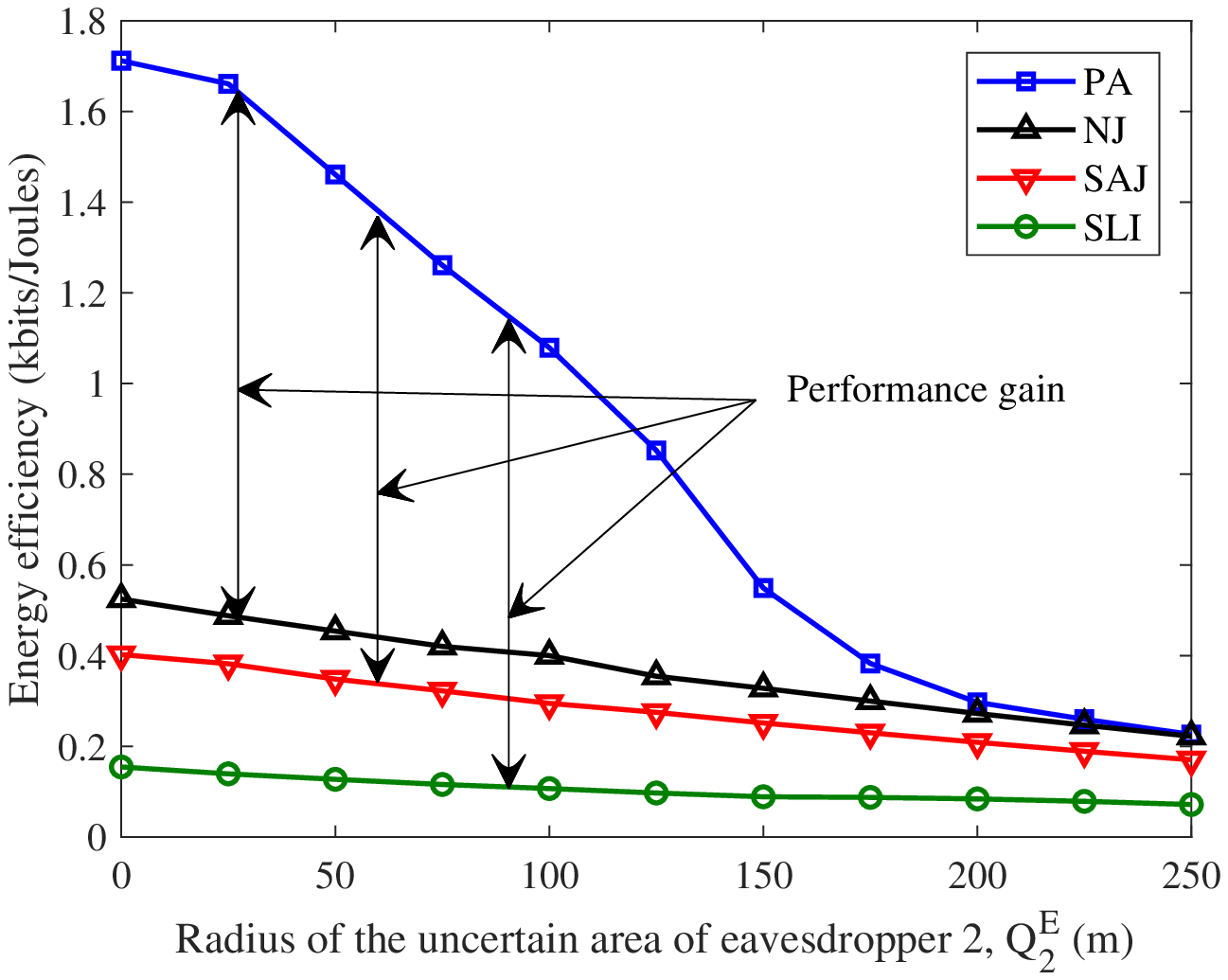} 
  \caption{Energy efficiency versus the radius of eavesdropper 2's uncertain area.}
  \label{EE_Vs_radius}
\end{figure}

Figure \ref{EE_Vs_P} shows the energy efficiency versus communication peak transmit power $P^{\mathrm I}_{\mathrm{peak}}$ for the ``PA'', ``NJ'', ``SAJ'', and ``SLI'' when the mission time duration $T$ is 50 s.
It can be observed that the energy efficiencies achieved by the ``PA'' and baseline schemes first increase with the communication peak transmit power budget.
This is due to the fact that increasing the communication transmit power budget can achieve a higher achievable data rate.
In particular, for low to moderate transmit power, the data rate gain due to a higher transmit power outweighs the cost of transmit power consumption leading to a rise in system energy efficiency.
However, the energy efficiency gain due to a higher values of $P^{\mathrm I}_{\mathrm{peak}}$ is diminishing and becomes saturated as the maximum system energy efficiency is achieved and the information UAV would clip the transmit power at the optimal value.
Moreover, the security constraint becomes more stringent for a larger $P^{\mathrm I}_{\mathrm{peak}}$ when the peak transmit power of artificial noise is fixed. As a result, to guarantee communication security, the information UAV may not always transmit with its full power in the high transmit power regime.
{Besides, it is observed that the system energy efficiency with perfect CSI is higher than that of ``PA''.
In fact, the CSI error arises from the uncertain area of eavesdroppers, which imposes a stringent information leakage constraints for the proposed scheme.
Therefore, more system resources are required to achieve secure communication.
As a result, the system performance degrades dramatically when there is an CSI error.
However, our proposed scheme can achieve the best performance among all the considered baseline schemes in the case of imperfect CSI.}
{Also, we can observe that the energy efficiency of ``ZAI'' is much lower than that of ``PA'' which presents the importance of variable UAV's flight speed for system performance.
In other words, varying the speed of UAV can help the system to exploit the system resources efficiently.}
Furthermore, we can observe that the increasing slope of ``PA'' is substantially higher than that of other baseline schemes.
In fact, the proper design of the artificial noise strategy of the multi-antenna jammer UAV offers the flexibility in designing the trajectory of information UAV and thus facilitates the efficient exploitation of power in our proposed scheme.

Figure \ref{EE_Vs_radius} depicts the energy efficiency of the considered system versus the radius of the uncertain area of potential eavesdropper 2 for the same schemes as in Figure \ref{EE_Vs_P}.
Note that we choose eavesdropper 2 instead of eavesdropper 1 in this figure.
The reason is that the uncertainty of eavesdropper 2 affects the trajectory of information UAV more significantly since its estimated location is on the straight locus from the initial location to the final location.
Although all schemes can guarantee communication security in all the considered cases, it can be observed that the energy efficiencies of both ``PA'' and baseline schemes decrease with the radius of uncertain areas.
Indeed, a larger eavesdropper's uncertain area imposes a more stringent security constraint on the system design, which reduces the flexibility in resource allocation leading to a lower system energy efficiency.
Furthermore, even with exact location information of eavesdroppers, all the three baseline schemes can only achieves a much smaller system energy efficiency compared to ``PA'', which again indicates the contribution of employing a multi-antenna jammer UAV and our proposed design.

\section{Conclusion}

In this paper, we jointly designed the information UAV's trajectory, the communication resource allocation strategy, and the jamming policy to maximize the system energy efficiency of a secure UAV-OFDMA communication system.
The joint design was formulated as a non-convex optimization problem taking into account the minimum data rate requirement, the maximum tolerable SINR leakage, the minimum safety distance between UAVs, and the imperfect location information of the potential eavesdroppers.
An iterative algorithm based on alternating optimization was proposed to achieve a suboptimal solution with a low computational complexity.
Simulation results illustrated that the proposed algorithm converges within a small number of iterations and demonstrated some interesting insights.
In particular,
(1) deploying a decided multiple-antenna UAV serves as a key to improve the system performance in both energy efficiency and communication security;
(2) employing a multi-antenna jammer UAV offers an enhanced flexibility in designing the trajectory of information UAV, which can combat the eavesdropper efficiently to improve the system energy efficiency;
(3) optimizing the trajectory of information UAV is important to improve the system energy efficiency.

\section{Appendix Proof of Theorem \ref{thm:rank_Z_leq_1}}

We follow a similar approach as in \cite{boshkovska2017robust} to prove Theorem 1.
First, it can be shown that the optimization problem \eqref{opt_prob:sub1_final} is jointly convex w.r.t. the optimization variables and satisfies the Slater's constraint qualification.
We first derive the Lagrangian function of 
\eqref{opt_prob:sub1_final}:
\begin{eqnarray} \label{eqn:L}
&& \mathfrak{L}({\bm Y},{\bm X},{\bm V},{\bm\mu},{\bm\nu},{\bm\vartheta},\mathcal{Z}) \\
&& = \sum_{n=1}^N \sum_{i=1}^{N_{\mathrm{F}}} \mathrm{Tr}\big(\mathbf{Z}_i^{\mathrm J}[n] (\mathbf{Y}_{i,n} - \mathbf{X}_{i,n} - \mathbf{V}_{i,n}) \big) \notag \\
&& - \sum_{n=1}^{\mathrm N} ( q_1^{(j^{\mathrm{A1}}_{\mathrm{in}})} + \mu_n + \nu_n \zeta^{\mathrm J}) \sum_{i=1}^{N_{\mathrm F}} \mathrm{Tr}(\mathbf{Z}_i^{\mathrm J}[n]) \notag \\
&& + \sum_{n=1}^N \sum_{e=1}^E \sum_{i=1}^{N_{\mathrm{F}}} \vartheta_{e,i,n} \underset{ \| \Delta \mathbf{t}_e^{\mathrm{E}}\| \leq Q_e^{\mathrm{E}} }{\min} \text{Tr}(\mathbf{H}_e^{\mathrm{JE}}[n] \mathbf{Z}_i^{\mathrm J}[n]) + \Delta, \notag
\end{eqnarray}
where $\Delta$ denotes the collection of terms that are not relevant for the proof.
Matrices $\mathbf{Y}_{i,n} \succeq \mathbf{0}, \forall i,n$, $\mathbf{X}_{i,n} \succeq \mathbf{0}, \forall i,n$, and $\mathbf{V}_{i,n} \succeq \mathbf{0}, \forall i,n$ are the Lagrange multiplier matrices for the constraint on matrix $\mathbf{Z}_i^{\mathrm J}[n]$ in $\mathrm{C3b}$, $\mathrm{C18}$, and $\mathrm{C19}$, respectively.
$\bm\mu=\{\mu_n,\forall n\}$, $\bm\nu=\{\nu_n,\forall n\}$, and $\bm\vartheta=\{\vartheta_{e,i,n},\forall e,i,n\}$ denote the Lagrange multipliers for constraints $\mathrm{C4b}$, $\mathrm{C5b}$, and $\mathrm{C7}$, respectively.
Considering \eqref{eqn:L}, the KKT conditions related to ${\mathbf{Z}_i^{\mathrm J}}^*[n]$ are given by
\begin{eqnarray}
\hspace*{-8mm}\mathbf{Y}_{i,n}^*, \mathbf{X}_{i,n}^*, \mathbf{V}_{i,n}^* &\succeq& \mathbf{0}, \, \mu_n^*,\nu_n^*,\vartheta_{e,i,n}^* \geq 0, \label{eqn:YXV_geq_0}\\
\hspace*{-8mm}{\mathbf{Z}_i^{\mathrm J}}^*[n] (\mathbf{Y}_{i,n}^* - \mathbf{X}_{i,n}^* - \mathbf{V}_{i,n}^*) &=& \mathbf{0}, \label{eqn:YZ_0}\\
\hspace*{-8mm}\nabla_{\mathbf{Z}}\mathfrak{L} &=& \mathbf{0}, \label{eqn:Nabla_Z}
\end{eqnarray}
where $\mathbf{Y}_{i,n}^*$, $\mathbf{X}_{i,n}^*$, $\mathbf{V}_{i,n}^*$, $\mu_n^*$, $\nu_n^*$, and $\vartheta_{e,i,n}^*$ are the optimal Lagrange multipliers for the dual problem of \eqref{opt_prob:sub1_final}.
Besides, \eqref{eqn:YZ_0} is the complementary slackness condition and is satisfied when the columns of ${\mathbf{Z}_i^{\mathrm J}}^*[n]$ lie in the null space of $\mathbf{Y}_{i,n}^* - \mathbf{X}_{i,n}^* - \mathbf{V}_{i,n}^*$.
To reveal the structure of $\mathbf{Z}_i^{\mathrm J}[n]$, we express the KKT condition in \eqref{eqn:Nabla_Z} as  
\begin{eqnarray} \label{eqn:Y_KKT}
\mathbf{Y}_{i,n}^*[n] &=& ( q_1^{(j^{\mathrm{A1}}_{\mathrm{in}})} + \mu_n + \nu_n \zeta^{\mathrm J}) \mathbf{I}_{N_{\mathrm J}} + \mathbf{X}_{i,n}^*[n] \notag \\
&+& \mathbf{V}_{i,n}^*[n] - \sum_{e=1}^E \vartheta_{e,i,n}^* \underset{ \| \Delta \mathbf{t}_e^{\mathrm{E}}\| \leq Q_e^{\mathrm{E}} }{\min} \mathbf{H}_e^{\mathrm{JE}}[n],
\end{eqnarray}
where $\underset{ \| \Delta \mathbf{t}_e^{\mathrm{E}}\| \leq Q_e^{\mathrm{E}} }{\min} \mathbf{H}_e^{\mathrm{JE}}[n]$ is a constant $N_\mathrm J \times N_\mathrm J$ matrix since we fix the jammer UAV's trajectory in this system.
For notation simplicity, we define $\mathbf{\Xi} = ( q_1^{(j^{\mathrm{A1}}_{\mathrm{in}})} + \mu_n + \nu_n \zeta^{\mathrm J}) \mathbf{I}_{N_{\mathrm J}} + \mathbf{X}_{i,n}^*[n] + \mathbf{V}_{i,n}^*[n] \succeq \mathbf{0}$ and $\mathbf{\Omega} = \sum_{e=1}^E \vartheta_{e,i,n}^* \underset{ \| \Delta \mathbf{t}_e^{\mathrm{E}}\| \leq Q_e^{\mathrm{E}} }{\min} \mathbf{H}_e^{\mathrm{JE}}[n] \succeq \mathbf{0}$.
From \eqref{eqn:YXV_geq_0}, since matrix $\mathbf{Y}_{i,n}^*[n] = \mathbf{\Xi} - \mathbf{\Omega}$ is positive semi-definite,  
\begin{eqnarray}
\lambda_{\Xi}^{\max} \geq \lambda_{\Omega}^{\max} \geq 0,
\end{eqnarray}
must hold, where $\lambda_{\Xi}^{\max}$ and $\lambda_{\Omega}^{\max}$ are the real-valued maximum eigenvalue of matrix $\mathbf{\Xi}$ and $\mathbf{\Omega}$, respectively.
Considering the KKT condition related to matrix ${\mathbf{Z}_i^{\mathrm J}}^*[n]$ in \eqref{eqn:YZ_0}, we can show that if $\lambda_{\Xi}^{\max} > \lambda_{\Omega}^{\max}$, matrix $\mathbf{Y}_{i,n}^*$ will become positive definite and full rank.
Besides, the maximum eigenvalue $\lambda_{\Xi}^{\max} > 0$ since $q_1^{(j^{\mathrm{A1}}_{\mathrm{in}})}$ is the energy-efficiency value of the system which is positive.
Thus, this would yield the solution ${\mathbf{Z}_i^{\mathrm J}}^*[n] = \mathbf{0}$.
On the other hand, if $\lambda_{\Xi}^{\max} = \lambda_{\Omega}^{\max}$, in order to have a bounded optimal dual solution, it follows that the null space of $\mathbf{Y}_{i,n}^*[n]$ is spanned by vector $\mathbf{u}_{\Omega,\max}$, which is the unit-norm eigenvector of $\Omega$ associated with eigenvalue $\lambda_{\Omega}^{\max}$.
As a result, we obtain the structure of the optimal energy matrix ${\mathbf{Z}_i^{\mathrm J}}^*[n]$ as  
\begin{eqnarray}
{\mathbf{Z}_i^{\mathrm J}}^*[n] = \delta \mathbf{u}_{\Omega,\max} \mathbf{u}_{\Omega,\max}^{\mathrm H}.
\end{eqnarray}
Therefore, $\text{Rank}({\mathbf{Z}_i^{\mathrm J}}^*[n]) \leq 1$.

%


\begin{IEEEbiography}[{\includegraphics[width=1in,height=1.4in]{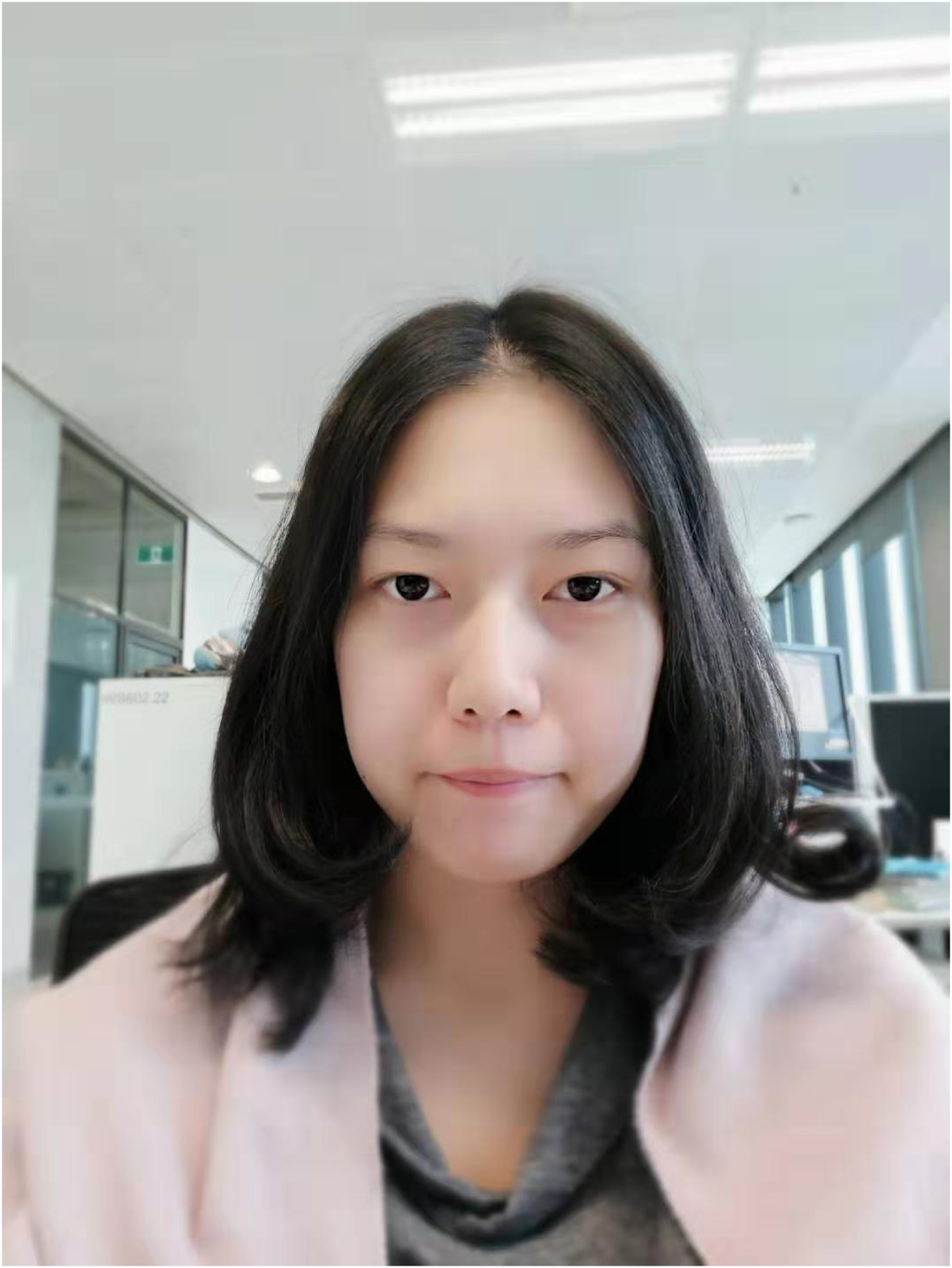}}]{Yuanxin Cai (S'19)} received the B.S. degree in Optical Information Science and Technology from the University of Electronic and Technology of China, Sichuan, China, in 2015 and the M.E. degree in Electrical Engineering and Telecommunication from the University of New South Wales, Sydney, Australia, in 2018. She is currently pursuing the Ph.D. degree in Telecommunication with the University of New South Wales, Sydney, Australia. Her current research interests include convex and non-convex optimization, UAV-assisted communication, resource allocation, physical-layer security, and green (energy-efficient) wireless communications.
\end{IEEEbiography}

\begin{IEEEbiography}[{\includegraphics[width=1in,height=1.4in]{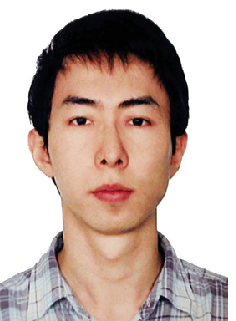}}]{Ruide Li} received the bachelor degree in flight vehicle propulsion engineering from Beijing Institute of Technology, Beijing, China, in 2011. He is currently pursuing the Ph.D. degree with the School of Information and Electronics, Beijing Institute of Technology, Beijing, China. From 2017 to 2019, he was a visiting student with the School of Electrical Engineering and Telecommunications, University of New South Wales, Sydney, Australia.
His research interests include unmanned aerial vehicle communications, resource allocation, and physical-layer security.
\end{IEEEbiography}

\begin{IEEEbiography}[{\includegraphics[width=1.1in,height=1.25in]{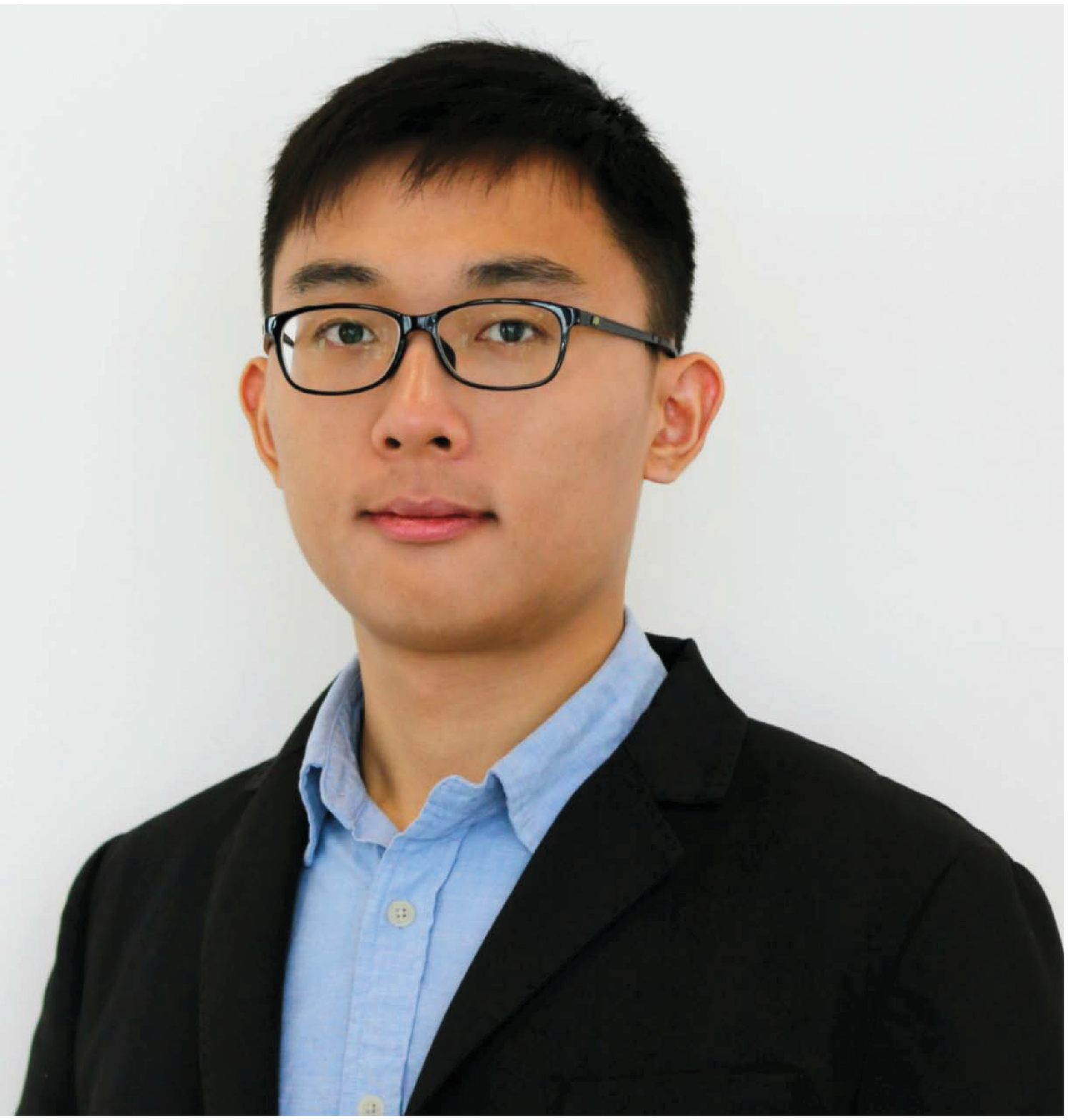}}]{Zhiqiang Wei (S'16-M'19)} received the B.E. degree in information engineering from the Northwestern Polytechnical University (NPU), Xi'an, China, in 2012 and the Ph.D. degree in Electrical Engineering and Telecommunications from the University of New South Wales, Sydney, Australia, in 2019. He is currently a postdoc research fellow in the University of New South Wales, Sydney, Australia. He received the Best Paper Awards at the IEEE International Conference on Communications (ICC), 2018.	His current research interests include statistic and array signal processing, non-orthogonal multiple access, millimeter wave communications, and resource allocation design.
\end{IEEEbiography}

\begin{IEEEbiography}[{\includegraphics[width=1in,height=1.25in,keepaspectratio]{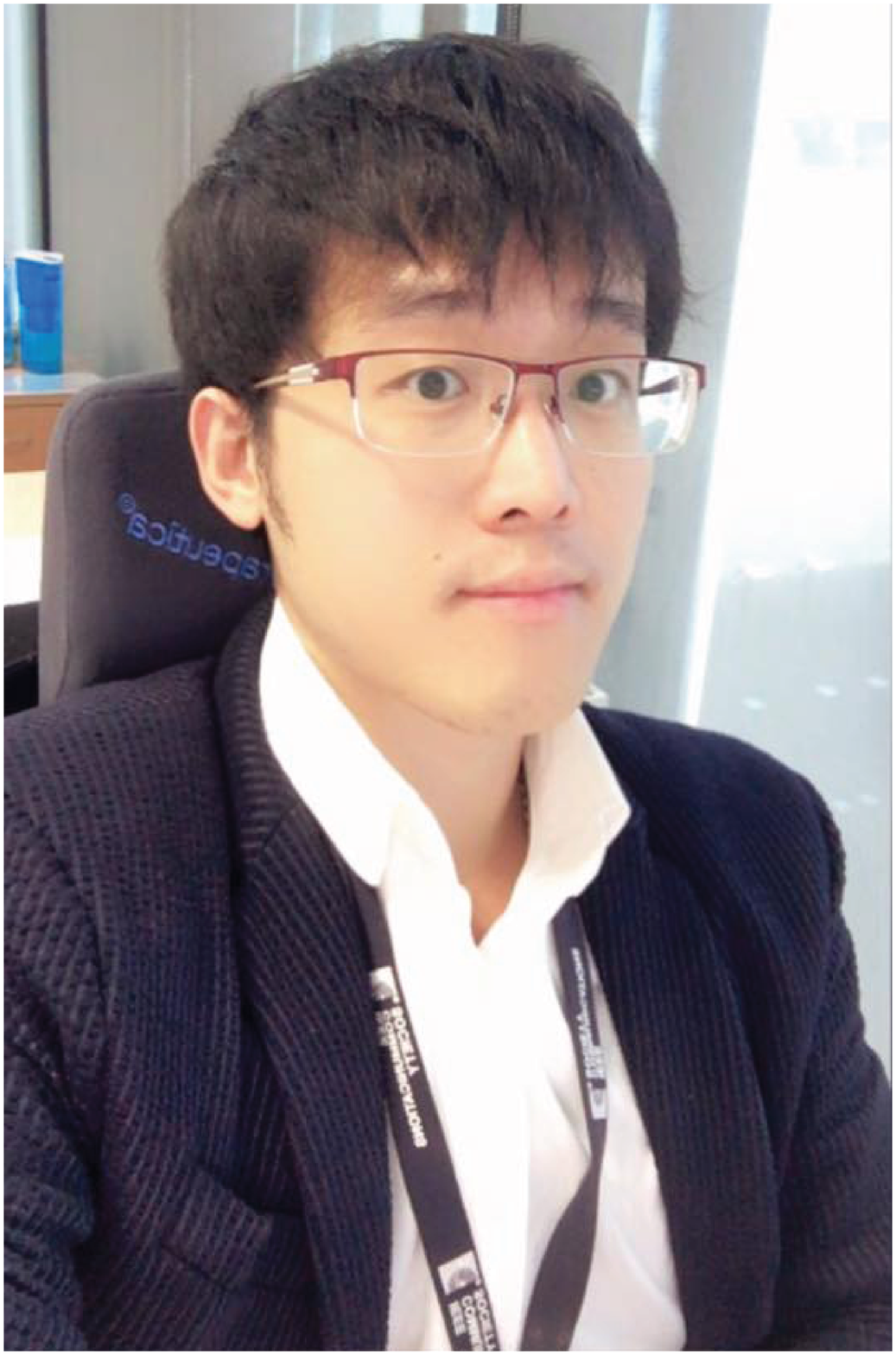}}]
{Derrick Wing Kwan Ng}(S'06-M'12-SM'17)  received the bachelor degree with first-class honors and the Master of Philosophy (M.Phil.) degree in electronic engineering from the Hong Kong University of Science and Technology (HKUST) in 2006 and 2008, respectively. He received his Ph.D. degree from the University of British Columbia (UBC) in 2012. He was a senior postdoctoral fellow at the Institute for Digital Communications, Friedrich-Alexander-University Erlangen-N\"urnberg (FAU), Germany. He is now working as a Senior Lecturer and a Scientia Fellow at the University of New South Wales, Sydney, Australia.  His research interests include convex and non-convex optimization, physical layer security, IRS-assisted communication, UAV-assisted communication, wireless information and power transfer, and green (energy-efficient) wireless communications.

Dr. Ng received the Best Paper Awards at the IEEE TCGCC Best Journal Paper Award 2018, INISCOM 2018, IEEE International Conference on Communications (ICC) 2018,  IEEE International Conference on Computing, Networking and Communications (ICNC) 2016,  IEEE Wireless Communications and Networking Conference (WCNC) 2012, the IEEE Global Telecommunication Conference (Globecom) 2011, and the IEEE Third International Conference on Communications and Networking in China 2008.  He served as an editorial assistant to the Editor-in-Chief of the \textsc{IEEE Transactions on Communications} from Jan. 2012 to Dec. 2019. He is now serving as an editor for the \textsc{IEEE Transactions on Communications},  the \textsc{IEEE Transactions on Wireless Communications}, and an area editor for the \textsc{IEEE Open Journal of the Communications Society}. Also, he was listed as a Highly Cited Researcher by Clarivate Analytics in 2018 and 2019.
\end{IEEEbiography}

\begin{IEEEbiography}[{\includegraphics[width=1in,height=1.4in]{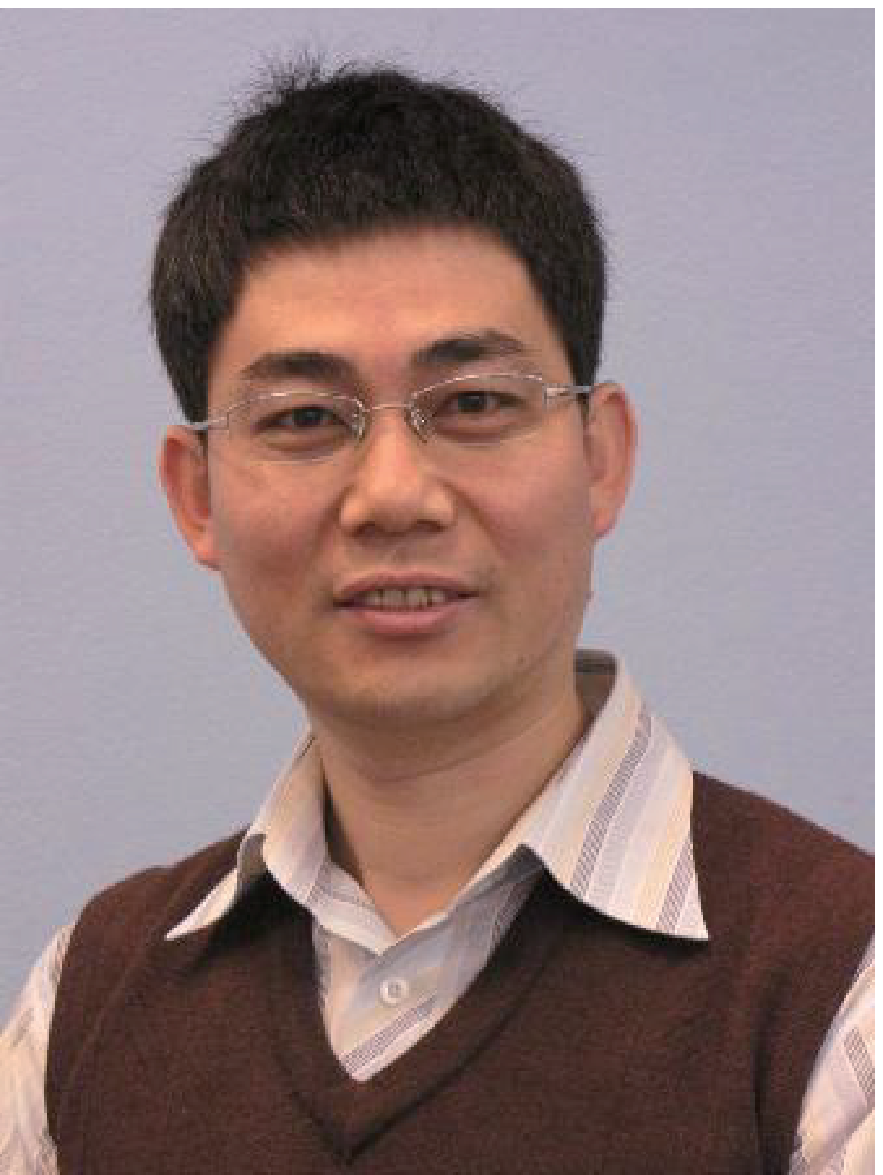}}]{Jinhong Yuan (M'02-SM'11-F'16)} received the B.E. and Ph.D. degrees in electronics engineering from the Beijing Institute of Technology, Beijing, China, in 1991 and 1997, respectively. From 1997 to 1999, he was a Research Fellow with the School of Electrical Engineering, University of Sydney, Sydney, Australia. In 2000, he joined the School of Electrical Engineering and Telecommunications, University of New South Wales, Sydney, Australia, where he is currently a Professor and Head of Telecommunication Group with the School. He has published two books, five book chapters, over 300 papers in telecommunications journals and conference proceedings, and 50 industrial reports. He is a co-inventor of one patent on MIMO systems and two patents on low-density-parity-check codes. He has co-authored four Best Paper Awards and one Best Poster Award, including the Best Paper Award from the IEEE International Conference on Communications, Kansas City, USA, in 2018, the Best Paper Award from IEEE Wireless Communications and Networking Conference, Cancun, Mexico, in 2011, and the Best Paper Award from the IEEE International Symposium on Wireless Communications Systems, Trondheim, Norway, in 2007. He is an IEEE Fellow and currently serving as an Associate Editor for the IEEE Transactions on Wireless Communications. He served as the IEEE NSW Chapter Chair of Joint Communications/Signal Processions/Ocean Engineering Chapter during 2011-2014 and served as an Associate Editor for the IEEE Transactions on Communications during 2012-2017. His current research interests include error control coding and information theory, communication theory, and wireless communications.
\end{IEEEbiography}

\end{document}